\documentclass[
				a4paper,fleqn,usenatbib,useAMS]{mnras}

\usepackage{newtxtext,newtxmath}

\usepackage[T1]{fontenc}
\usepackage{ae,aecompl}

\usepackage{graphicx}	
\usepackage{amsmath}	
\usepackage{amssymb}	

\usepackage[version=3]{mhchem}
\usepackage{datetime}
\usepackage{multirow}
\usepackage{booktabs}
\usepackage{float}
\usepackage{gensymb}
\usepackage{wasysym}
\usepackage{amsfonts}
\usepackage{courier}
\usepackage{caption}
\usepackage{subcaption}
\usepackage{multicol}   
\usepackage{pdflscape}	
\usepackage[para]{threeparttable}
\usepackage{blindtext}
\usepackage{enumitem}

\newdateformat{mydate}{\twodigit{\THEDAY}{ }\monthname[\THEMONTH], \THEYEAR}
\mydate
\newdate{reportdate}{11}{10}{2017}
\newdate{datadate}{02}{2}{2017}

\newcommand{\rom}[1]{%
  \textup{\uppercase\expandafter{\romannumeral#1}}%
}
\newcommand*\mean[1]{\overline{#1}}

\title[Properties of the single Jovian planet population]{Properties of the single Jovian planet population and the pursuit of Solar system analogues}

\author[M. T. Agnew et al.]{
Matthew T. Agnew,$^{1}$ 
Sarah T. Maddison,$^{1}$ and
Jonathan Horner$^{2}$
\\
$^{1}$Centre for Astrophysics and Supercomputing, Swinburne University of Technology, Hawthorn, Victoria 3122, Australia\\
$^{2}$University of Southern Queensland, Toowoomba, Queensland 4350, Australia\\
}

\date{Accepted 2018 April 2. Received 2018 March 15; in original form 2018 January 17}

\pubyear{2017}

\begin{document}
\label{firstpage}
\pagerange{\pageref{firstpage}--\pageref{lastpage}}
\maketitle

\begin{abstract}
While the number of exoplanets discovered continues to increase at a rapid rate, we are still to discover any system that truly resembles the Solar system. Existing and near future surveys will likely continue this trend of rapid discovery. To see if these systems are Solar system analogues, we will need to efficiently allocate resources to carry out intensive follow-up observations. We seek to uncover the properties and trends across systems that indicate how much of the habitable zone is stable in each system to provide focus for planet hunters. We study the dynamics of all known single Jovian planetary systems, to assess the dynamical stability of the habitable zone around their host stars. We perform a suite of simulations of all systems where the Jovian planet will interact gravitationally with the habitable zone, and broadly classify these systems. Besides the system's mass ratio ($M_{pl}/M_{star}$), and the Jovian planet's semi-major axis  ($a_{pl}$) and eccentricity ($e_{pl}$), we find that there are no underlying system properties which are observable that indicate the potential for planets to survive within the system's habitable zone. We use $M_{pl}/M_{star}$, $a_{pl}$ and $e_{pl}$ to generate a parameter space over which the unstable systems cluster, thus allowing us to predict which systems to exclude from future observational or numerical searches for habitable exoplanets. We also provide a candidate list of 20 systems that have completely stable habitable zones and Jovian planets orbiting beyond the habitable zone as potential first order Solar system analogues.
\end{abstract}

\begin{keywords}
methods: numerical -- planets and satellites: dynamical evolution and stability -- planets and satellites: general -- planetary systems -- astrobiology
\end{keywords}



\section{Introduction}
\label{sec:introduction}
A key goal of exoplanetary science is to find Earth analogue planets - planets that might have the right conditions for life to both exist and be detectable. Given that the only location we know of that hosts life is the Earth, that search is strongly biased towards looking for planetary systems that strongly resemble our own - Solar system analogues. While we have seen an explosion of exoplanet discoveries in the last decade that is sure to continue \citep[e.g.][]{Sullivan2015,Dressing2017}, the discovery of Solar system analogues still proves to be a decidedly challenging goal. While Jupiter-sized planets have been detected for over 20 years, these are often very close to their host stars \citep[e.g.][]{Mayor1995,Charbonneau2000}. It has only been much more recently that we have begun to detect Jupiter-sized planets on decade long orbital periods; the  so-called Jupiter analogues \citep{Boisse2012,Wittenmyer2014,Wittenmyer2016,Kipping2016,Rowan2016}. Similarly, discoveries of lower mass planets have become more common \citep{Wright2015,Vogt2015,Gillon2017}, thanks in a large part to the Kepler survey \citep{Borucki2011,Morton2016}. The current count of confirmed exoplanets now exceeds 3500\footnote{As of 18 January 2018 (NASA Exoplanet Archive, exoplanetarchive.ipac.caltech.edu).}. As a result, we can begin to consider the exoplanet population as a whole in order to better understand any overarching properties of the sample, and to also provide a means to exclude existing systems from further follow up in our search for Solar system analogues.

Due to the observational biases inherent to the radial velocity (RV) method \citep{Wittenmyer2011a,Dumusque2012b}, a great deal of work has gone into attempting to theoretically predict where additional exoplanets could remain stable in existing systems, via both predictions of regions of stability and/or instability \citep{Jones2001,Jones2002,Jones2005,Jones2010,Giuppone2013} and dynamical simulations \citep{Rivera2007,Thilliez2014,Kane2015,Thilliez2016}. The large size of Jovian planets means they are often easier to detect and can dominate RV signals. For this reason, it has been suggested that the seeming abundance of single Jovian planet systems is the result of an observational bias rather than a true reflection of the exoplanet population \citep{Marcy2005a,Cumming2008}. 

In the Solar system, Jupiter is thought to have played an integral role in determining the Solar system architecture that we see today \citep[e.g.][]{Gomes2005b,Horner2009,Walsh2011,Izidoro2013,Raymond2014,Brasser2016,Deienno2016}. A number of authors have investigated the role Jupiter may have played in nurturing the right environment on Earth for life to have prospered \citep[e.g.][]{Bond2010,Carter-Bond2012,Bond2010b,Martin2013a,Quintana2014,OBrien2014} although this is still an active area of research for which debate continues \citep[e.g.][]{Horner2008,Horner1908,Horner2010,Horner2012,Horner2013,Grazier2016}. However, without a clear answer, the search for Solar system analogues and the search for habitable exoplanets remains tightly coupled. Finding a true Solar system analogue is inherently challenging due to the small transit and radial velocity signals of the inner rocky planets, and the large decade to century long orbits of the outer giant planets. Because of this, we begin with a simplified definition of a Solar system analogue, that being: a Sun-like star with a rocky planet in the habitable zone (HZ), and a Jovian planet orbiting beyond the outer boundary of the HZ. Thus, searching for single Jovian systems that are capable of hosting hidden Earth-like planets in the HZ becomes a natural starting point in the search for habitable exoplanets and Solar system analogues. 


\cite{Agnew2017} took a sample of single Jovian planet systems and used N-body simulations to produce a candidate list of systems that could host a $1\ \textrm{M}_{\oplus}$ planet on a stable orbit within the system's habitable zone, and that could be detected with current or near-future instruments. Here we expand upon that earlier work by examining all currently known single Jovian planet systems to (1) identify any overarching trends (that may be the result of formation or evolution scenarios) within the single Jovian planet population, (2) exclude planetary architectures within which the system's HZ would be unstable, and (3) provide a candidate list to guide future observing efforts in the search for Solar system analogues. 

In section \ref{sec:method}, we describe the method used to calculate the boundaries of the HZ, how we select the single Jovian planet systems which we wish to simulate, and detail the numerical simulations used to dynamically analyse these systems. We then discuss our results in section \ref{sec:results}, and present a candidate list of Solar system analogues for use by future planet hunters. We summarise our findings in section \ref{sec:summary}.

\section{Method}
\label{sec:method}
We first consider the existing single Jovian population from the NASA Exoplanet Archive\footnote{Our sample of single Jovian planets was obtained from the NASA Exoplanet Archive, exoplanetarchive.ipac.caltech.edu, on 27 March 2017 which gave an initial sample of 771 planets.}, removing all systems with incomplete stellar or planetary properties. We then calculate the HZ boundaries for each system using the method outlined in \cite{Kopparapu2014}. This allows us to estimate those systems for which the HZ will likely be stable (due to the distance of the Jovian planet from the HZ). For those systems for which the Jovian is located close enough to the HZ to potentially perturb the region, we then peform dynamical simulations to ascertain the degree to which this occurs.

For our analysis of the sample of the single Jovian planet population, we accept the stellar and planetary properties as they are presented in the relevant databases, acknowledging that there may be uncertainties associated with these parameters.

\subsection{Habitable Zone}
\label{subsec:hz}
We calculate the HZ boundaries using the method outlined by \cite{Kopparapu2014}, which is only valid for stars with $2600$~K $\leq T_\mathrm{eff} \leq 7200$~K. They present an equation for the astrocentric distance of different regimes for the inner and outer boundary of the HZ as
\begin{align}
	d_{\textrm{HZ}} = \sqrt{\frac{L/L_\odot}{S_{\textrm{eff}}}} \rm{au},
\end{align}
where $L$ is the luminosity of the star, and $S_{\textrm{eff}}$ is calculated as
\begin{align}
	S_{\textrm{eff}} = S_{\textrm{eff}\odot} + aT_\star + bT_\star^2+cT_\star^3+dT_\star^4,
\end{align}
where $T_\star = T_{\textrm{eff}} - 5780$ K, and $a$, $b$, $c$, $d$ and $S_{\textrm{eff}\odot}$ are constants depending on the planetary mass considered, $M_\textrm{pl}$, and the HZ boundary regime being used. Here, we assume a $1\ \textrm{M}_{\oplus}$ planet, and use a conservative HZ boundary regime utilising the Runaway Greenhouse boundary for the inner edge, and the Maximum Greenhouse boundary for the outer edge \citep{Kopparapu2014}. This corresponds with the constants shown in Table~\ref{tab:HZ_params}. \cite{Kane2014} found that these boundaries are significantly influenced by uncertainties of the stellar parameters, but we use the best fit values as presented in the NASA Exoplanet Archive.

\begin{table}
	\caption{The constants used to calculate the HZ for our simulations, assuming the Earth-like planet we are searching for is $1\ \textrm{M}_{\oplus}$, as presented in \protect\cite{Kopparapu2014}}
   	\label{tab:HZ_params}
   	\centering
	\begin{tabular}{c c c c c}
    	\cmidrule(r){3-3}  \cmidrule(r){5-5}  
    								& 	&{Runaway Greenhouse}  & & {Maximum Greenhouse }\\

    	\cmidrule(r){1-1} \cmidrule(r){3-3}  \cmidrule(r){5-5} 	 
        $a$							&	& 	$1.332\times10^{-4}$& 	&	$6.171\times10^{-5}$\\
        $b$							&	& 	$1.58\times10^{-8}$& 	&	$1.698\times10^{-9}$\\
        $c$							&	& 	$-8.308\times10^{-12}$& 	&	$-3.198\times10^{-12}$\\
        $d$							&	& 	$-1.931\times10^{-15}$& 	&	$-5.575\times10^{-16}$\\
        $S_{\textrm{eff}\odot}$		&	& 	$1.107$& 				&	$0.356$\\
    	\cmidrule(r){1-1} \cmidrule(r){3-3}  \cmidrule(r){5-5} 
   	\end{tabular}
\end{table}

\subsection{System Selection}
\label{subsec:predicting_stable_regions}
\begin{figure*}
	\centering
	\includegraphics[width=\linewidth]{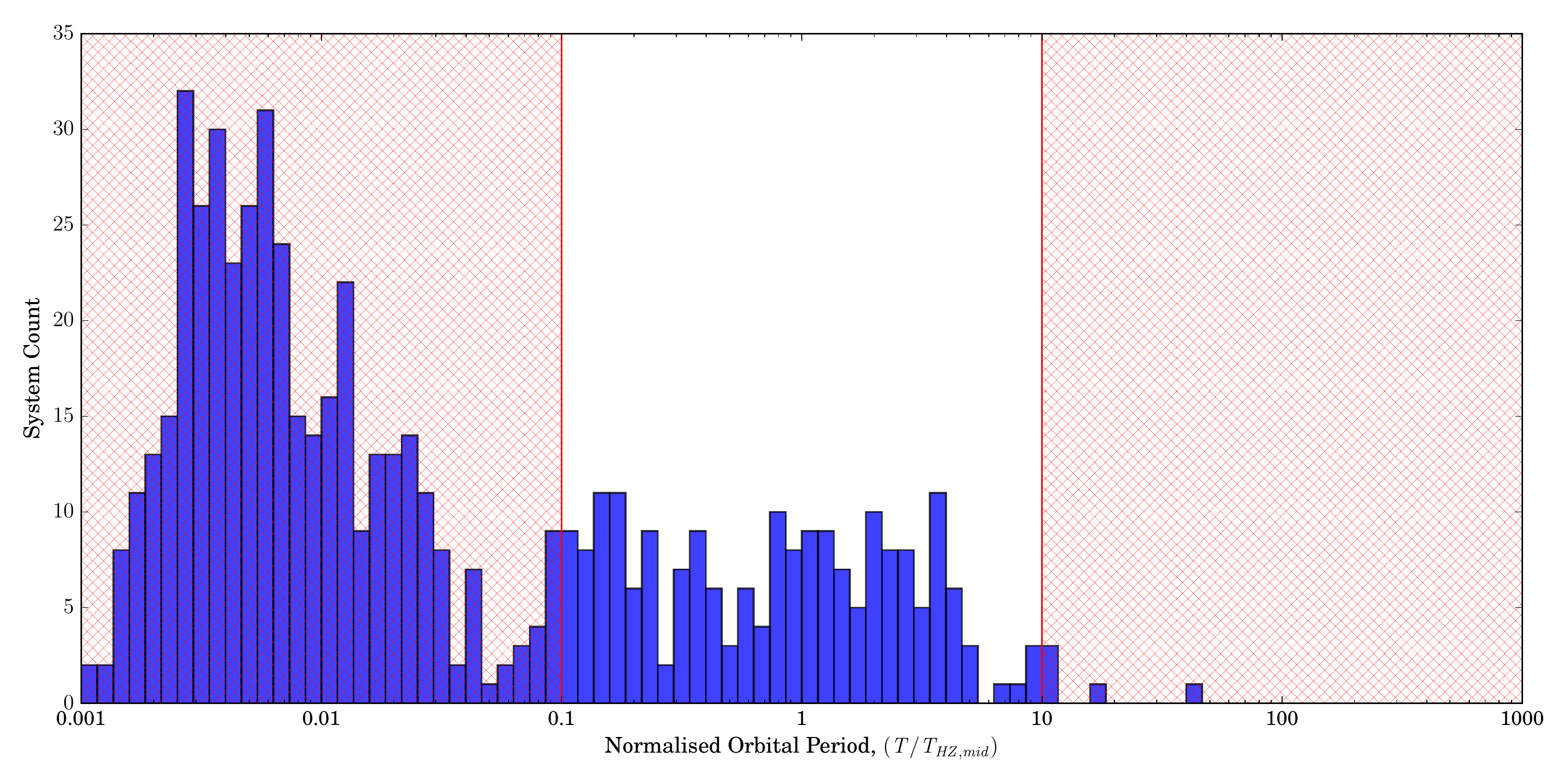}
	\caption{All 542 currently known single Jovian planet systems that have the required stellar and planetary parameters, and that satisfy the $2600$~K $\leq T_\mathrm{eff} \leq 7200$~K criterion for the HZ calculation. The $x$-axis is the orbital period of each Jovian planet, normalised by the the period in the centre of the HZ. The red, hashed area represents the cut-off for systems that do not satisfy the $0.1\ T_{\textrm{HZ}} \leq T_{\textrm{Jovian}} \leq 10\ T_{\textrm{HZ}}$ criterion.}
	\label{fig:selection}
\end{figure*}

The single Jovian planet population as of 27 March 2017 is made up of 771 systems\footnote{Those systems with only one planet that has $R_{\mathrm{pl}}>6$ Earth radii or $M_{\mathrm{pl}}>50$ Earth masses in lieu of available radius data.}. We  remove from this sample those systems that are missing planetary or stellar properties, which excludes 175 systems. From the 596 that remain, 54 systems feature stellar temperatures that fall outside the range $2600$~K $\leq T_\mathrm{eff} \leq 7200$~K required by the \cite{Kopparapu2014} HZ calculation, and so these too are removed from our sample. This yields the final sample of 542 systems.

For all systems for which the Jovian planet is greater than 10 Hill radii from the midpoint of the HZ, we expect little to no gravitational stirring within the HZ \citep{Jones2005,Jones2010,Giuppone2013}. In such systems, computational resources are wasted on simulating completely stable HZs. Of the 542 systems in our sample, a total of 360 systems fell into this category, of which 355 had an interior Jovian, and 5 an exterior Jovian. We excluse these systems from our suite of simulations, and simply tag them as having wholly stable HZs. In our sample, all of the systems remaining that are expected to gravitationally stir the HZ have a Jovian planet with an orbital period $0.1\ T_{\textrm{HZ,mid}} \leq T_{\textrm{Jovian}} \leq 10\ T_{\textrm{HZ,mid}}$, and so we use this criteria as a slightly more conservative cut than $a_{\mathrm{pl}} > 10 R_{\textrm{Hill}}$. A histogram illustrating the 542 systems with the orbital period cuts overlaid can be seen in Figure~\ref{fig:selection}.

Using the Jovian orbital period criterion outlined above leaves a total of 182 single Jovian systems that could, potentially, exhibit a degree of instability within the HZ. In this work, we simulate this sample to investigate the impact of the Jovian planets on the stability across the HZ.

\subsection{Dynamical Simulations}
\label{subsec:dynamical_analysis}

In order to assess whether a system with a known Jovian planet could host an Earth-like world in its habitable zone, we carry out a suite of detailed $N$-body simulations. We distribute a large number of massless test particles (TPs) through the HZ of the systems in which we are interested, and integrate the evolution of their orbits forwards in time for a period of 10 million years. This is a computationally intensive endeavour, and the simulations we present below required a total of six months of continuous integration across the several hundred computing cores available to us. In order to facilitate a timely analysis, in this work we solely examine the scenario of co-planar systems - in which the orbits of our putative exo-Earths are always set to move in the same plane as the Jovian planet. This focus on co-planar orbits is common in exoplanetary science, being the standard assumption in the modelling of the orbits of newly discovered multiple exoplanet systems \citep[e.g.][]{Robertson2012,Horner2013a,Horner2014}, where no information is currently held on the mutual inclination between the orbits of the known planets. In some studies \citep[e.g.][]{Horner2011,Wittenmyer2012,Wittenmyer2013b}, we have shown that mutual inclination between exoplanet orbits typically acts to render a system less stable. As such the assumption of co-planarity is a mechanism by which we maximise the potential for a given system to exhibit a dynamically stable habitable zone.

Both our own Solar system and those multiple exoplanet systems have demonstrably very low mutual inclinations \citep{Lissauer2011,Lissauer2011a,Fang2012,Figueira2012,Fabrycky2014}. However, we \textcolor{blue}{acknowledge that those systems} are not perfectly flat. Once a small amount of mutual inclination is added to a previously co-planar system, it opens up the possibility for the excitation of both the inclination and eccentricity of objects that would otherwise have been moving on mutually stable orbits. As such, in the future, we plan to expand this work to investigate the impact of small, but non-zero, mutual inclinations on the stability (or otherwise) of those systems for which this work predicts a dynamically stable outcome. In this work, our focus remains on the study of the most optimistic scenario, that being perfect co-planarity, the aforementioned caveat must be kept in mind.

To test the dynamic stability of each system, we use the \textsc{swift} $N$-body software package \citep{Levison1994} to run a series of simulations with massless TPs. We randomly distribute 5000 TPs throughout the HZ of each system within the ranges shown in Table~\ref{tab:tp_params}. The upper bound of the TP eccentricity of 0.3 was selected as a reasonable upper value for an orbit to remained confined to the HZ \citep{Jones2005}. While a planet  may be considered habitable at eccentricities as high as $0.5 < e < 0.7$ depending on the response time of the atmosphere-ocean system \citep{Williams2002,Jones2005}, we are interested in planets and systems that more closely resemble the Earth and the Solar system.

The simulation of each system was run for a total integration time of $T_{\textrm{sim}} = 10^7$ years, or until all TPs were removed from the system. TPs ejected beyond an astrocentric distance of $250$~au are removed from the simulations. The time step of each simulation was calculated to be $1/50$ of the smallest initial orbital period of the TPs, or the Jovian planet if it was interior to the HZ. 

\begin{table}
	\caption{The range of orbital parameters within which the test particles were randomly distributed throughout the HZ.}
   	\label{tab:tp_params}
   	\centering
	\begin{tabular}{c c c c}
        \cmidrule(r){3-4}                    
        						&	& Min					& Max \\
        \cmidrule(r){1-2} \cmidrule(r){3-4} 
        $a$ (au)				&	& HZ$_{\mathrm{min}}$	& HZ$_{\mathrm{max}}$\\
        $e$					&	& 0.0					& 0.3 \\
        $i$ ($\degree$)		&	& 0.0					& 0.0 \\
        $\Omega$ ($\degree$)	&	& 0.0					& 0.0 \\
        $\omega$ ($\degree$)	&	& 0.0					& 360.0 \\
        $M$ ($\degree$)		&	& 0.0					& 360.0 \\
        \cmidrule(r){1-2} \cmidrule(r){3-4}
   	\end{tabular}
\end{table}
\section{Results \& Discussion}
\label{sec:results}
For each of the 182 systems simulated, the lifetime of each TP and the resulting number of TPs that survived was recorded, as well as their initial semi-major axis and eccentricity. The resulting stability of individual systems, as well as the entire population, can then be analysed.

\subsection{Dynamical Classifications}

\begin{figure*}
\centering
	\begin{subfigure}{0.31\textwidth}
  		\centering
  		\includegraphics[width=\textwidth]{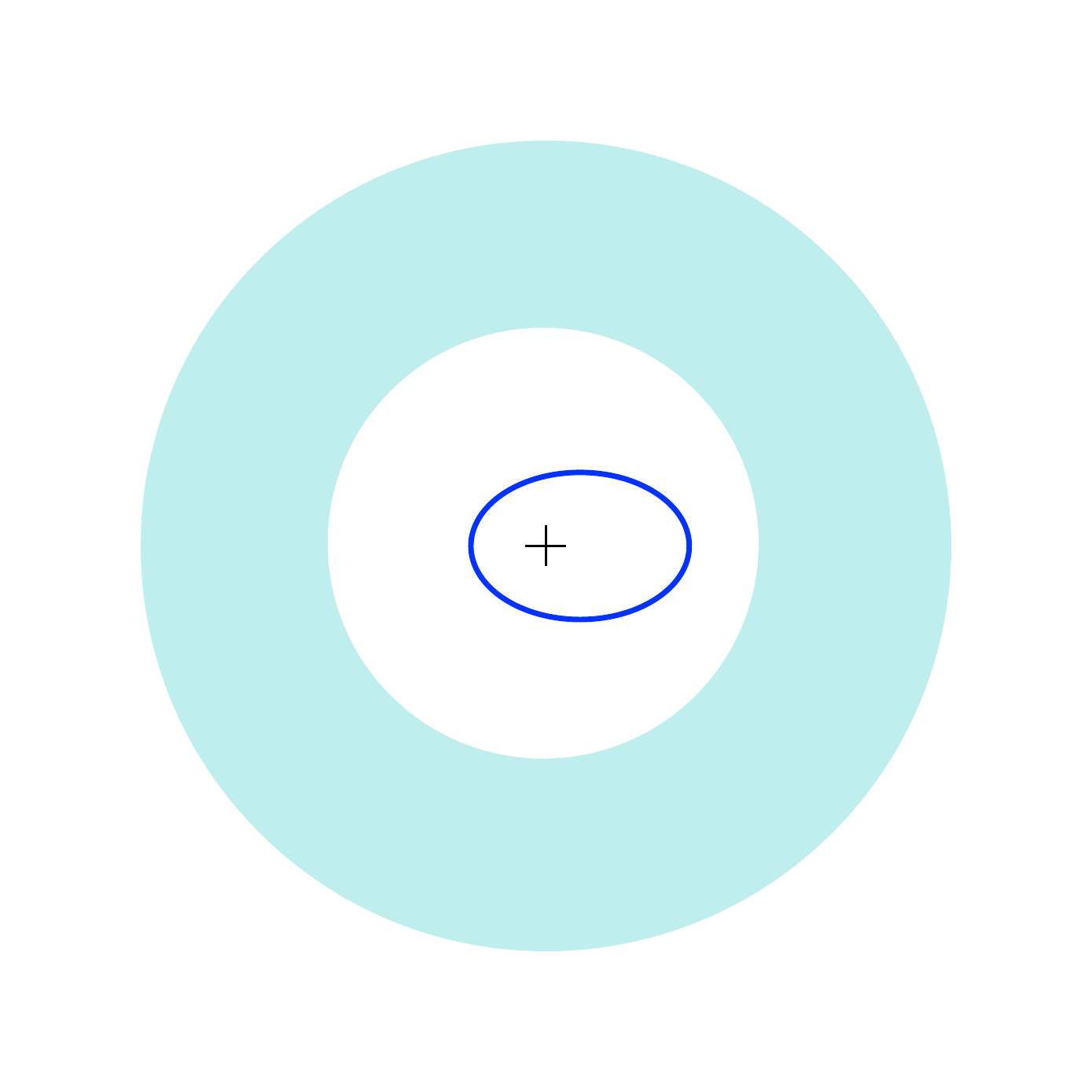}
  		\caption{\rom{1}-a Interior}
	\end{subfigure}
	\begin{subfigure}{0.31\textwidth}
  		\centering
  		\includegraphics[width=\textwidth]{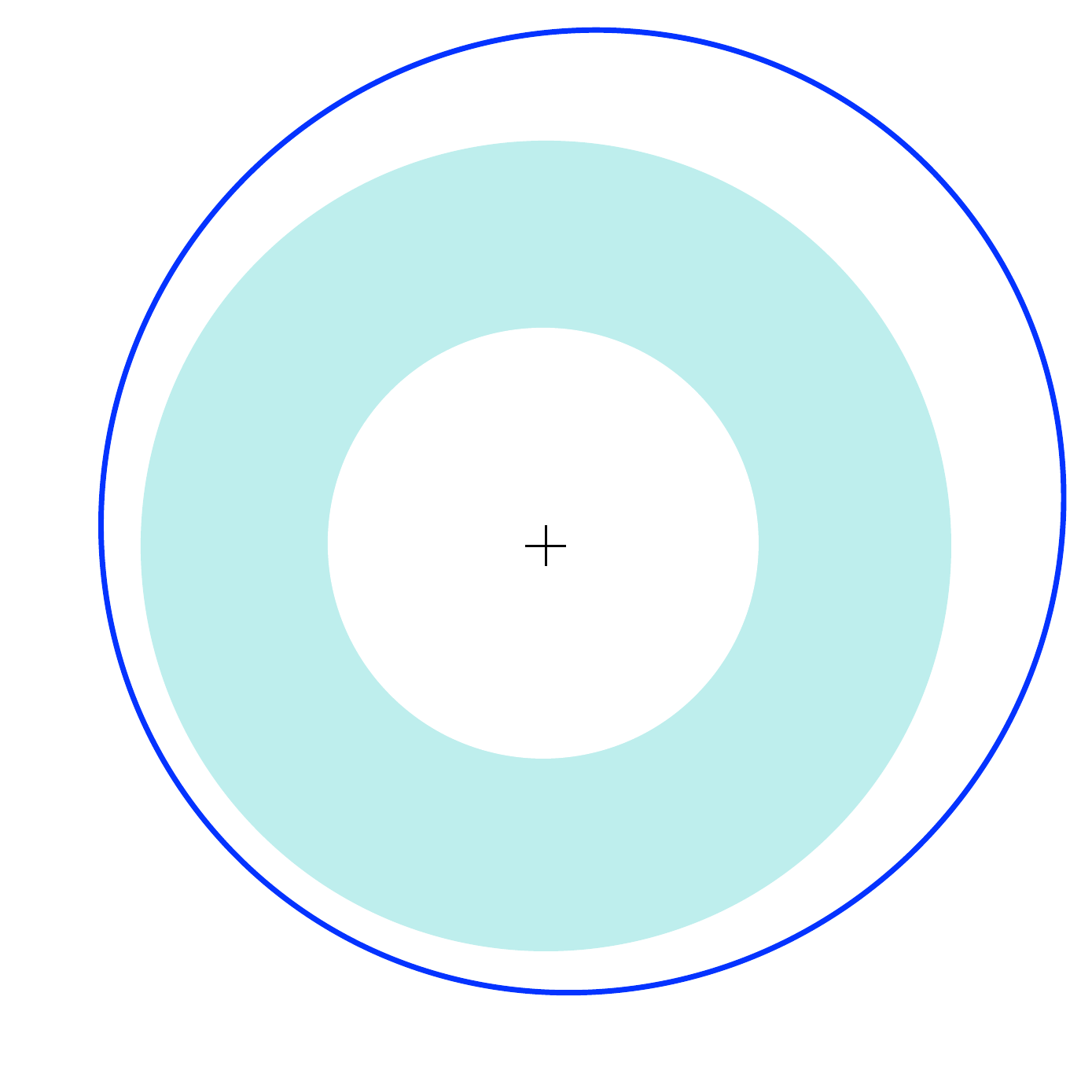}
  		\caption{\rom{2}-a Exterior}
	\end{subfigure}
	\begin{subfigure}{0.31\textwidth}
  		\centering
  		\includegraphics[width=\textwidth]{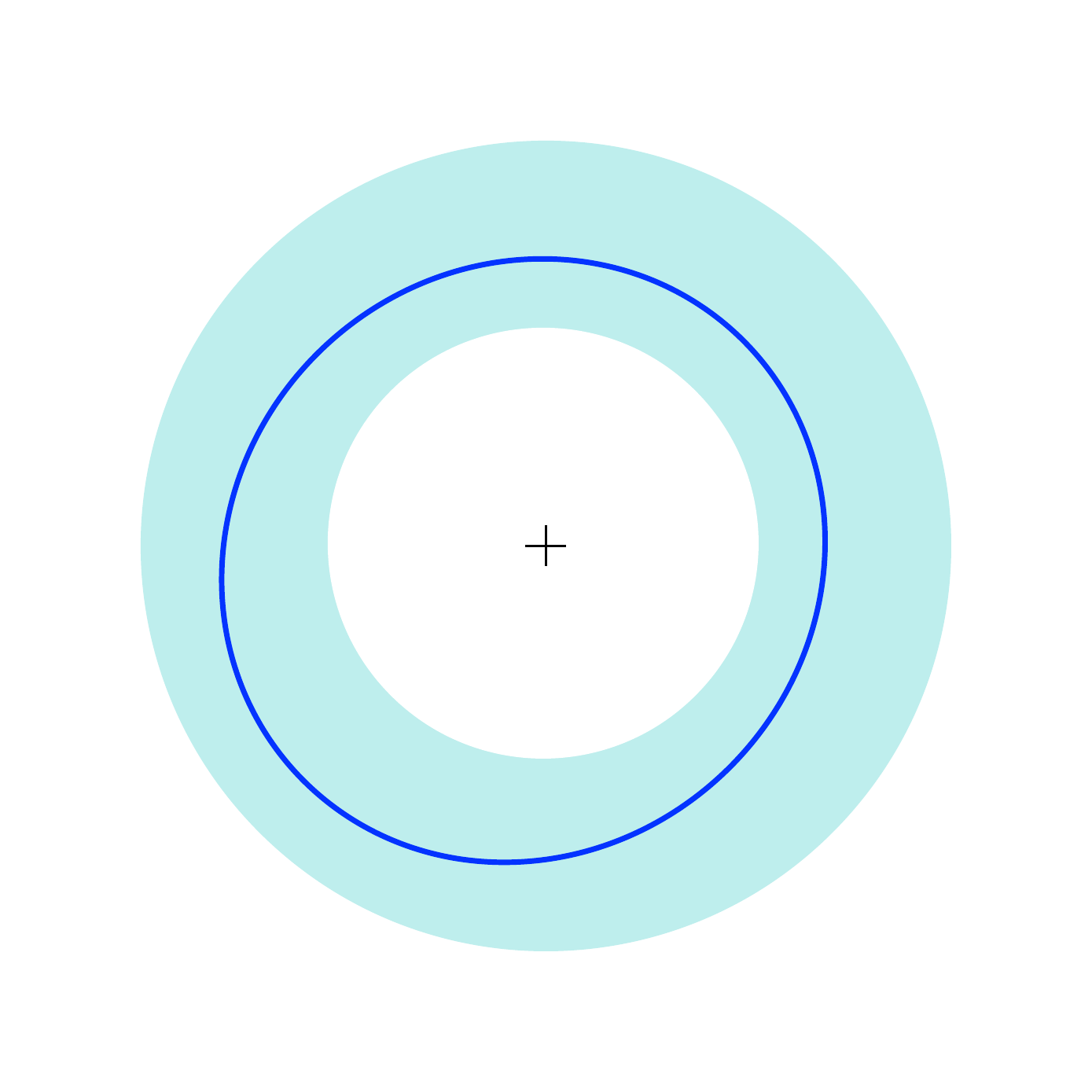}
  		\caption{\rom{3} Embedded}
	\end{subfigure}
	
	\begin{subfigure}{0.31\textwidth}
  		\centering
  		\includegraphics[width=\textwidth]{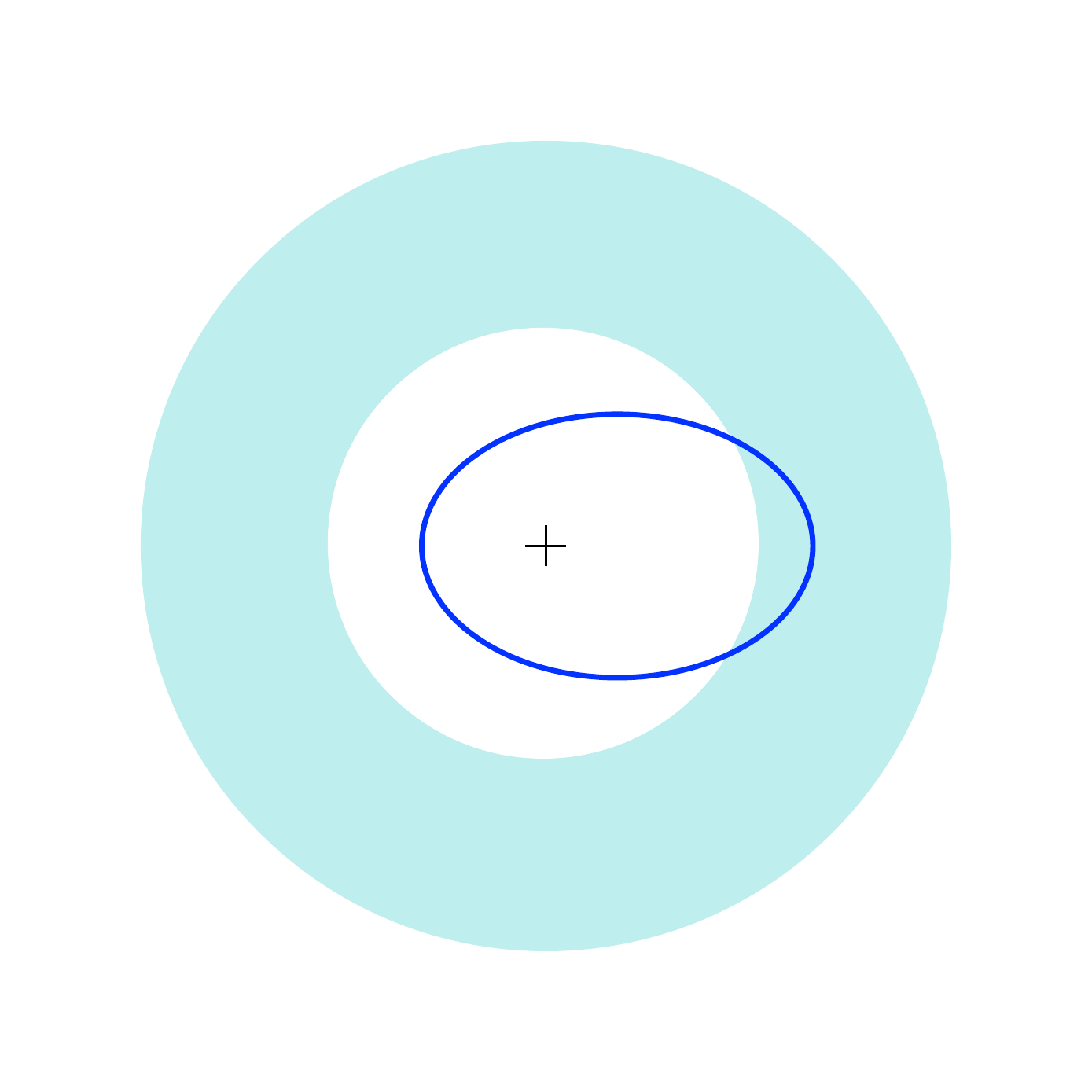}
  		\caption{\rom{1}-b Interior \& touching}
	\end{subfigure}
	\begin{subfigure}{0.31\textwidth}
  		\centering
  		\includegraphics[width=\textwidth]{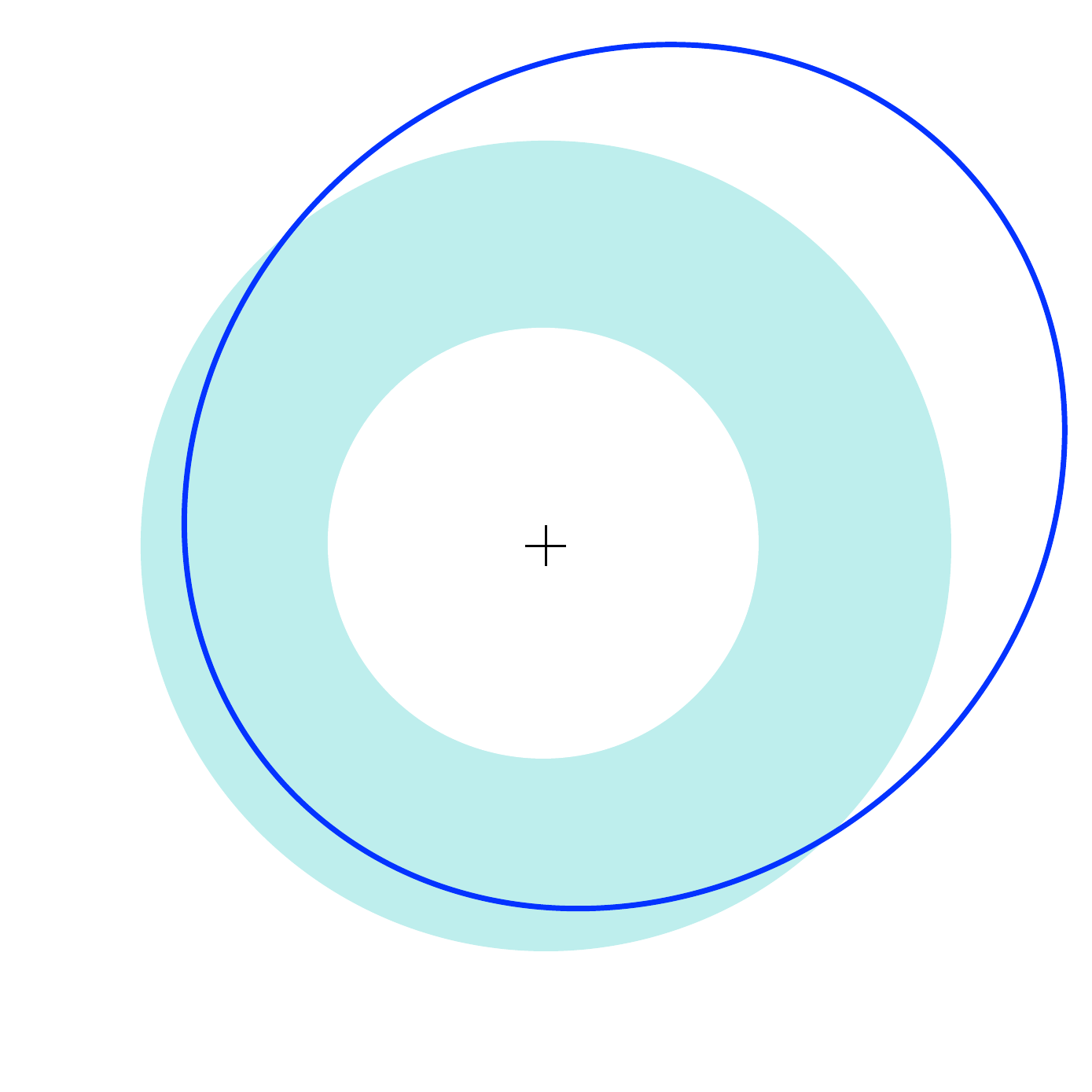}
  		\caption{\rom{2}-b Exterior \& touching}
	\end{subfigure}
	\begin{subfigure}{0.31\textwidth}
  		\centering
  		\includegraphics[width=\textwidth]{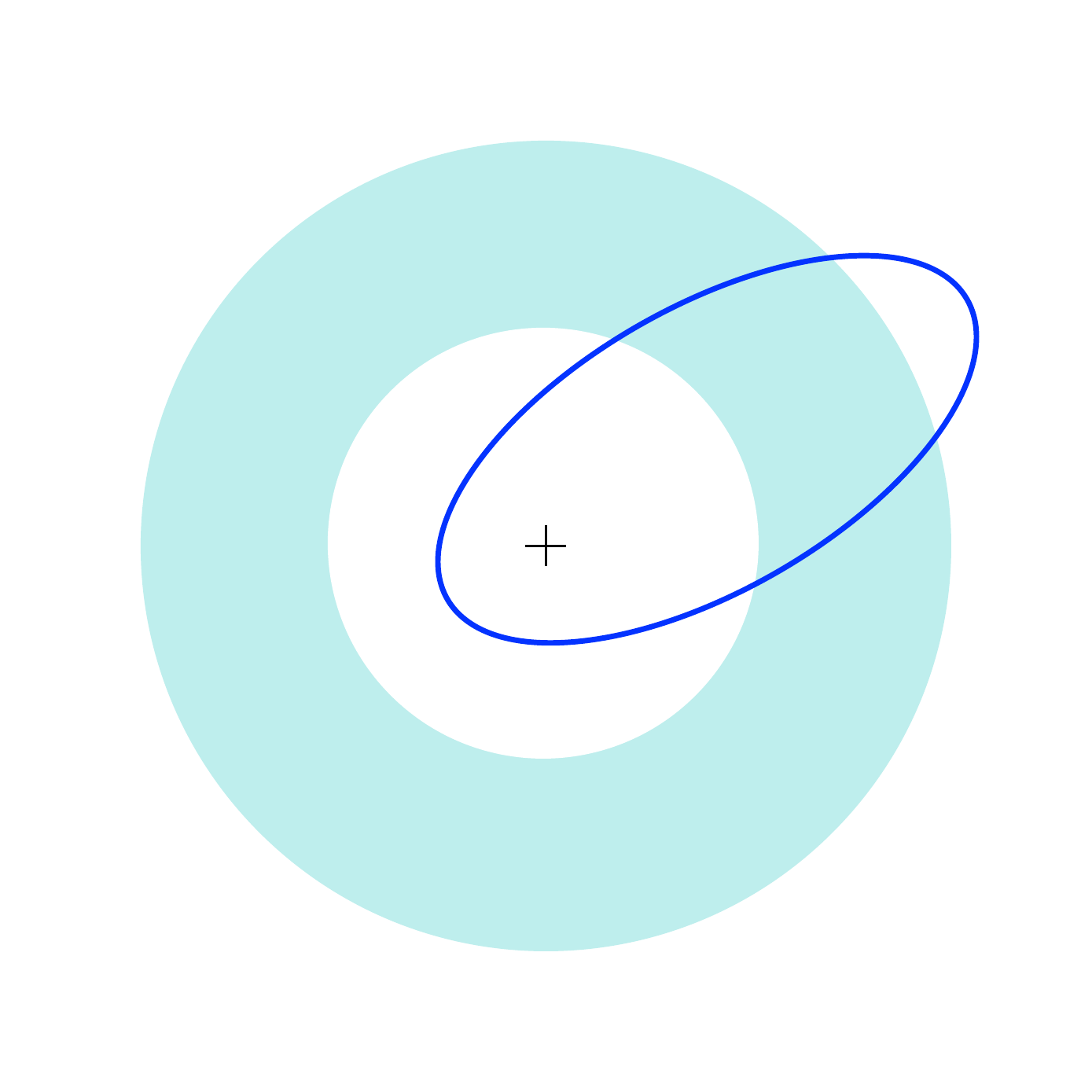}
  		\caption{\rom{4} Traversing}
	\end{subfigure}
	\caption{The six dynamical classes of single Jovian planet systems: (a) \rom{1}-a Interior, (b) \rom{2}-a Exterior, (c) \rom{3} Embedded, (d) \rom{1}-b Interior and touching, (e) \rom{2}-b Exterior and touching, and (f) \rom{4} Traversing. The green annulus represents the HZ of the system, while the blue ellipse represents the orbit of the Jovian planet. The black cross represents the star.}
	\label{fig:class}
\end{figure*}

Our 182 systems are broadly be divided into 6 dynamical classifications based on the apsides of the Jovian planet relative to the boundaries of the HZ. These classes are: 
\begin{description}
\item [\rom{1}-a] Interior,
\item [\rom{1}-b] Interior \& touching,
\item [\rom{2}-a] Exterior,
\item [\rom{2}-b] Exterior \& touching,
\item [\rom{3}] Embedded, and
\item [\rom{4}] Traversing.
\end{description}
The locations of the Jovian planet's periastron, $r_{\mathrm{peri}}$, and apastron, $r_{\mathrm{ap}}$, relative the HZ are listed in Table~\ref{tab:class_desc}.
\begin{table}
	\caption{The location of the Jovian planet's periastron, $r_{\mathrm{peri}}$, and apastron, $r_{\mathrm{ap}}$, relative to the HZ for each dynamical classification.}
   	\label{tab:class_desc}
   	\centering
	\begin{tabular}{l c c c}
        \toprule               
        Class		& $r < a_{\mathrm{HZ,in}}$	&	$a_{\mathrm{HZ,in}} < r < a_{\mathrm{HZ,out}}$	& $a_{\mathrm{HZ,out}} < r$ \\
        \toprule
        \rom{1}-a 	& $r_{\mathrm{peri}}$, $r_{\mathrm{ap}}$ & -											&  - \\
        \rom{1}-b 	& $r_{\mathrm{peri}}$ 		& $r_{\mathrm{ap}}$ 										&  -\\
        \midrule
        \rom{2}-a 	& -							& -														& $r_{\mathrm{peri}}$, $r_{\mathrm{ap}}$\\
        \rom{2}-b 	& -							& $r_{\mathrm{peri}}$									& $r_{\mathrm{ap}}$ \\
        \midrule
        \rom{3}		& -	 						& $r_{\mathrm{peri}}$, $r_{\mathrm{ap}}$				& - \\
        \midrule
        \rom{4} 	& $r_{\mathrm{peri}}$ 		& -														& $r_{\mathrm{ap}}$ \\
        \toprule
   	\end{tabular}
\end{table}
These are demonstrated in the schematic in Figure~\ref{fig:class}. The number of single Jovian systems in each dynamical class is listed in Table~\ref{tab:class}.
\begin{table}
	\caption{The number of single Jovian systems within each dynamical class.}
   	\label{tab:class}
   	\centering
	\begin{tabular}{l c c c}
        \toprule               
        	Class					&	& Sub-count					& Total \\
        \midrule
        \rom{1}-a Interior	&	& 64					& \multirow{2}{*}{88}  \\
        \rom{1}-b Interior \& Touching	&	& 24					&  \\
        \midrule
        \rom{2}-a Exterior	&	& 34					& \multirow{2}{*}{60}  \\
        \rom{2}-b Exterior \& Touching	&	& 26					&  \\
        \midrule
        \rom{3}	Embedded	&	& 					& 11 \\
        \midrule
        \rom{4}	Traversing& 	&  					& 23 \\
        \midrule
        Total & & & 182 \\
        \toprule
   	\end{tabular}
\end{table}

\begin{figure*}
	\centering
	\includegraphics[width=\linewidth]{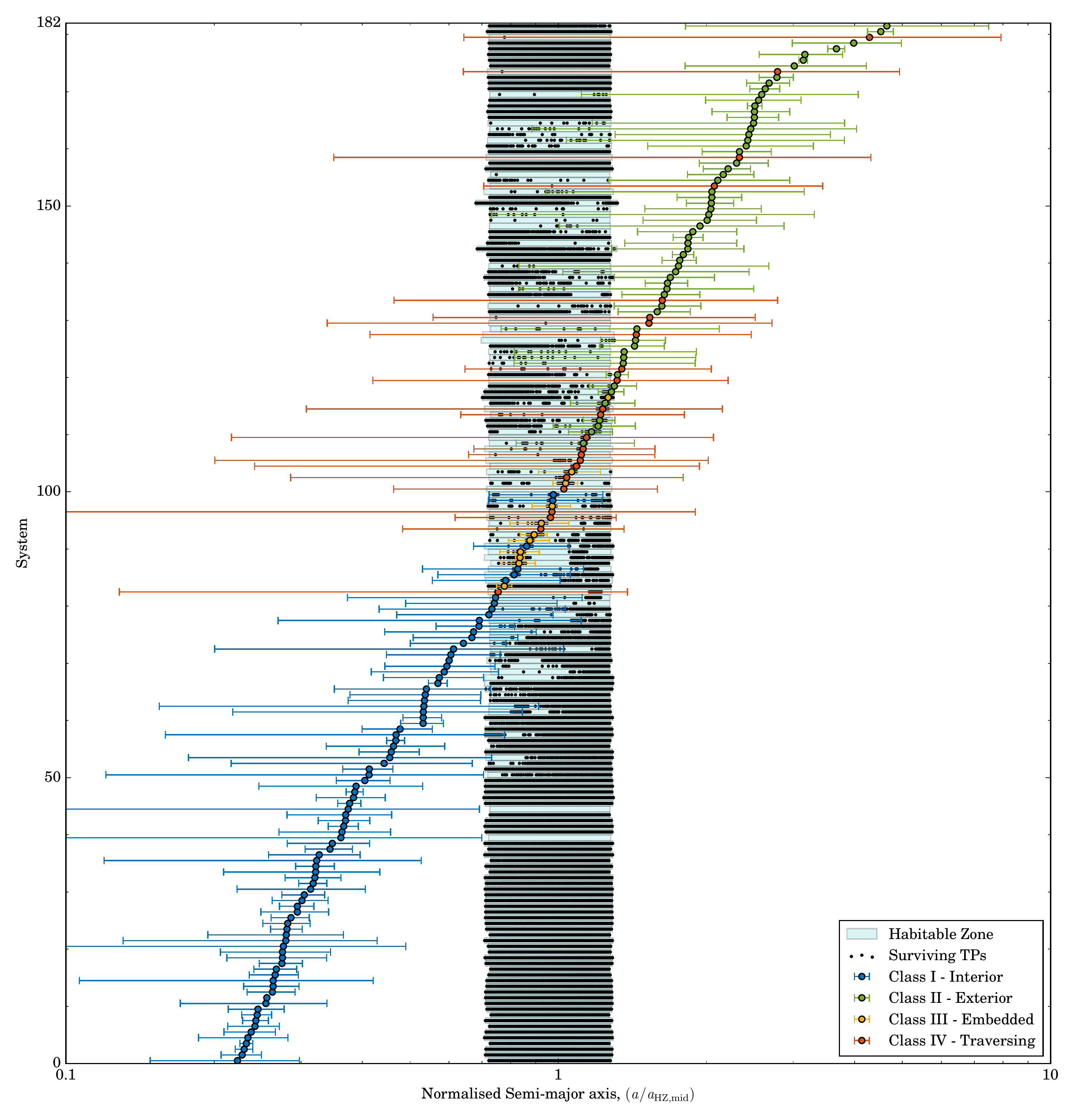}
	\caption{All 182 systems simulated plotted on a normalised semi-major axis ($a/a_\mathrm{HZ,mid}$) $x$-axis. The coloured points represent the Jovian planet and error bars the apsides of its orbit. The green region represents the HZ, and the black points are those TPs of the initial 5000 that are still surviving at the end of the simulation.}
	\label{fig:all_sim_syst}
\end{figure*}

In Figure~\ref{fig:all_sim_syst} we show the entire population of single Jovian planet systems simulated. Figure~\ref{fig:all_sim_syst} very clearly demonstrates the importance of $e_\mathrm{pl}$ on HZ stability where those systems with high eccentricities -- represented by large error bars -- have far less stable HZs. Class \rom{1} and \rom{2} systems have large regions of stability within the HZ, and this decreases with the Jovian planet's increasing eccentricity as the system begins to approach becoming a class \rom{4} (traversing) system.
We would intuitively expect class \rom{3} and \rom{4} systems to be unstable. However, we do see TPs surviving for the duration of the simulation in some cases, and thus we will investigate these classes in greater detail.

\begin{figure}
	\centering
	\includegraphics[width=\linewidth]{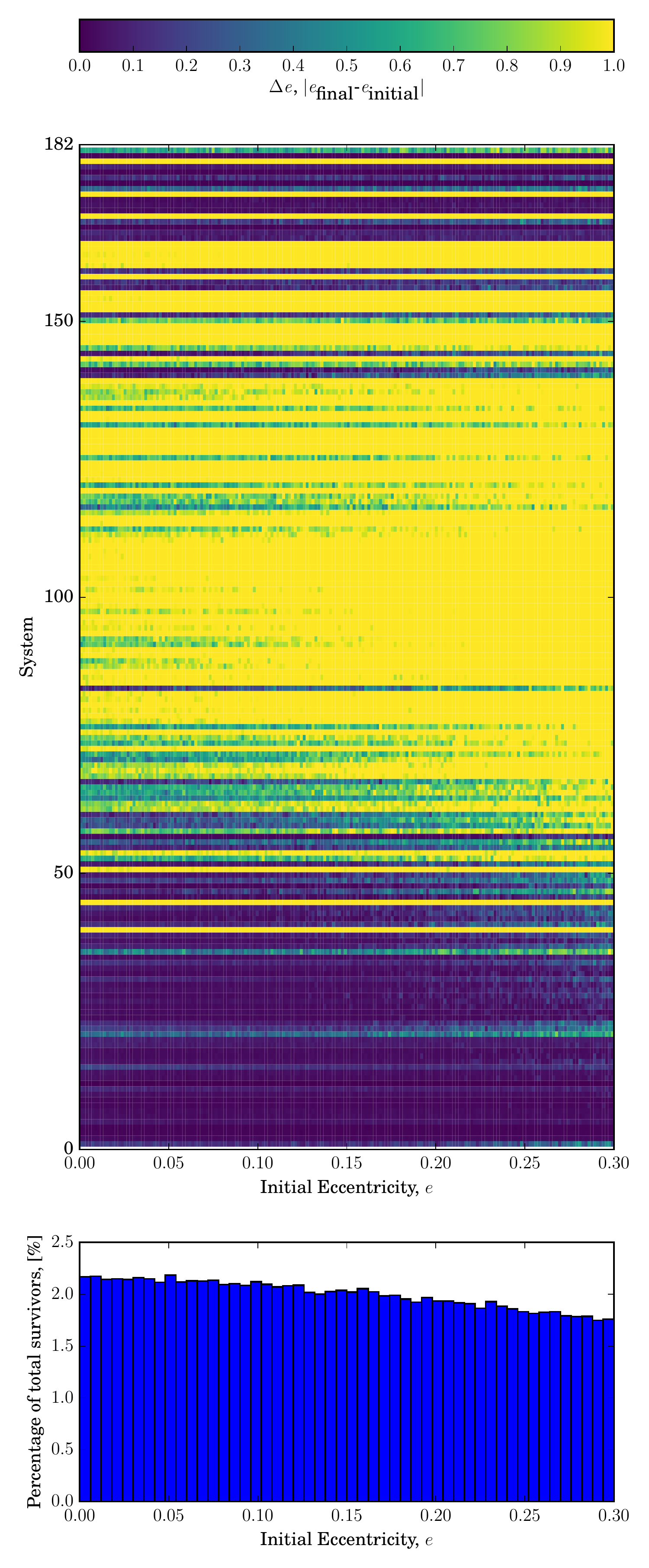}
	\caption{{The change in TP eccentricities for all 182 systems simulated. We plot  the initial TP eccentricity against the systems ordered along the $y$-axis as in Figure~\ref{fig:all_sim_syst}. The colour represents the mean change in eccentricity of all TPs within each bin ($\mean{\Delta e} = 1/n \sum{\Delta e_n}$). The lower panel shows a 1D histogram of the percentage of all surviving TPs that particular initial eccentricity values make up.}}
	\label{fig:all_e_syst}
\end{figure}

\begin{figure}
	\centering
	\includegraphics[width=\linewidth]{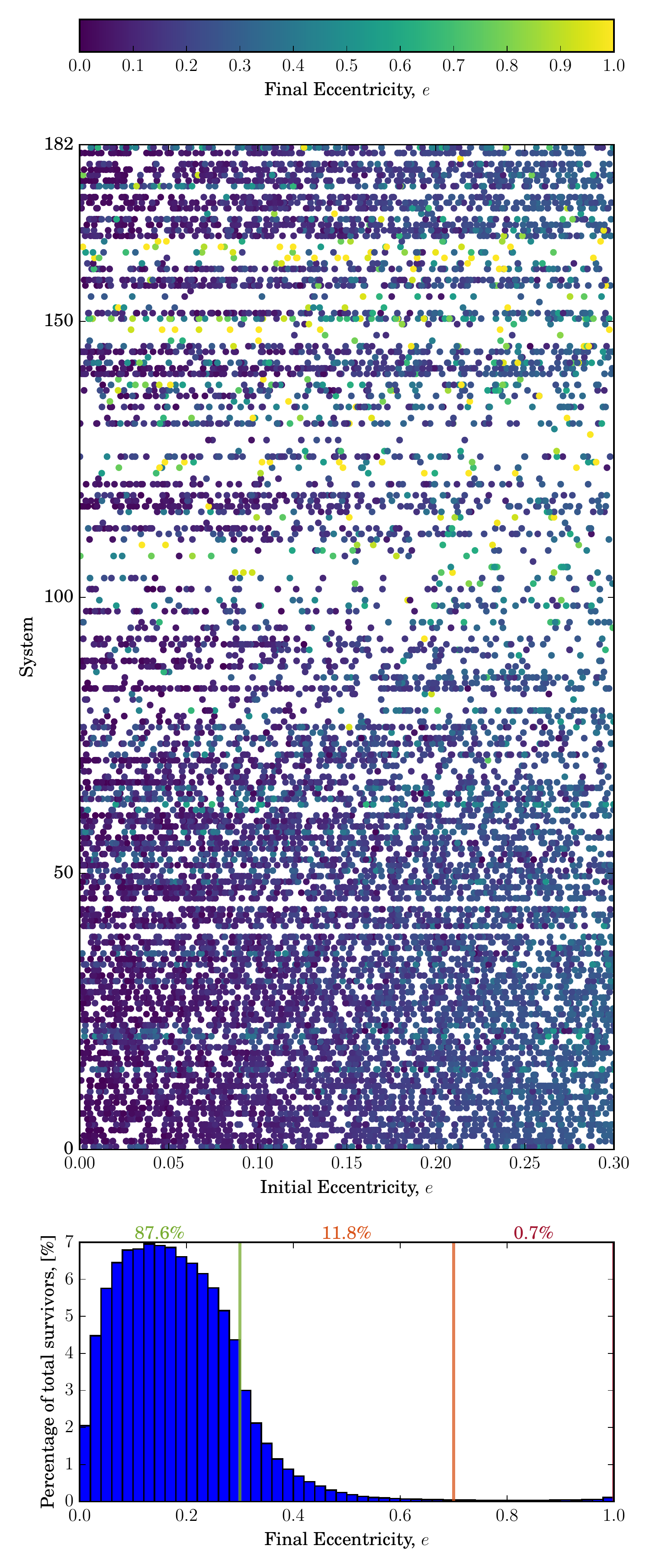}
	\caption{{Similar to Figure~\ref{fig:all_e_syst}, but colour now represents  final eccentricities of the TP in all 182 systems. The lower panel shows the 1D histogram of the percentage of all surviving TPs that have a particular final eccentricity. The various cuts for habitability for an $e<0.3$ and $e<0.7$ are overlaid in green and orange respectively, while the percentage of surviving TPs that fall within these cuts are listed above.}}
	\label{fig:all_e_syst_fin}
\end{figure}

In Figures~\ref{fig:all_e_syst} and \ref{fig:all_e_syst_fin} we again show the entire population of single Jovian planet systems simulated, but this time considering how the orbital eccentricities of the stable TPs evolve. 
Figure~\ref{fig:all_e_syst} demonstrates the level of eccentricity excitation of each TP, shown by the change in eccentricity ($\Delta e$) of the simulated TPs, where a change of 1 is used to indicate complete removal of the TP from the system. As there are 5000 TPs per simulation, we bin the TPs in 182 equally spaced bins, and take the mean eccentricity change of all TPs within each bin ($\mean{\Delta e} = 1/n \sum{\Delta e_n}$) so as to not lose information from excessive stacking of points. We plot the systems in the same order as they appear in Figure~\ref{fig:all_sim_syst} to allow comparison. Unsurprisingly, TPs with higher initial eccentricities are removed in systems where the Jovian planet is located near the HZ, as the apsides of their orbits mean they begin to experience close encounters with the Jovian planet while lower eccentricity TPs may not. In the 1D histogram in lower panel of Figure~\ref{fig:all_e_syst}, we combine the initial eccentricity data of the survivors across all 182 simulated systems. This more clearly highlights that, over the entire population, lower initial eccentricity TPs are more likely to survive.

Figure~\ref{fig:all_e_syst_fin} shows the final eccentricities of the surviving TPs against their initial eccentricity for all systems. We see that the majority of the final eccentricities of  surviving TPs are  low ($e < 0.3$), suggesting little eccentricity excitation. However, there are clear examples of higher eccentricity TPs surviving, shown by the green and yellow points scattered across the plot. We combine the final eccentricity data of all the survivors across the 182 systems in the 1D histogram in the lower panel. This clearly shows that TPs with final eccentricities less than 0.3 ($87.6\%$) dominate the surviving TP population. Combined with the level of excitation shown in the top panel, this demonstrates that surviving TPs experience low levels of  excitation, and that when TPs are excited they tend to be removed entirely.

In the search for an Earth-like planet, we focus specifically on TPs that have a final eccentricity of less than 0.3, as this is a value that generally leads to a HZ confined orbit \citep{Jones2005}. However, studies suggest that a planet may still receive sufficient luminosity to be considered habitable with eccentricities as high as 0.7, depending on a range of planet properties \citep{Williams2002,Jones2005}. Taking the cut off of $e<0.3$ indicates that $87.6\%$ of the surviving TPs would be considered potentially habitable, based solely on this criterion, while the more optimistic cut of $e<0.7$ takes that total up to $99.3\%$.

\subsubsection{Class~III HZ--embedded Jovian planets}
\label{subsubsec:mmr}
\begin{figure}
	\begin{subfigure}{\linewidth}
    		\centering
    		\includegraphics[width=\linewidth]{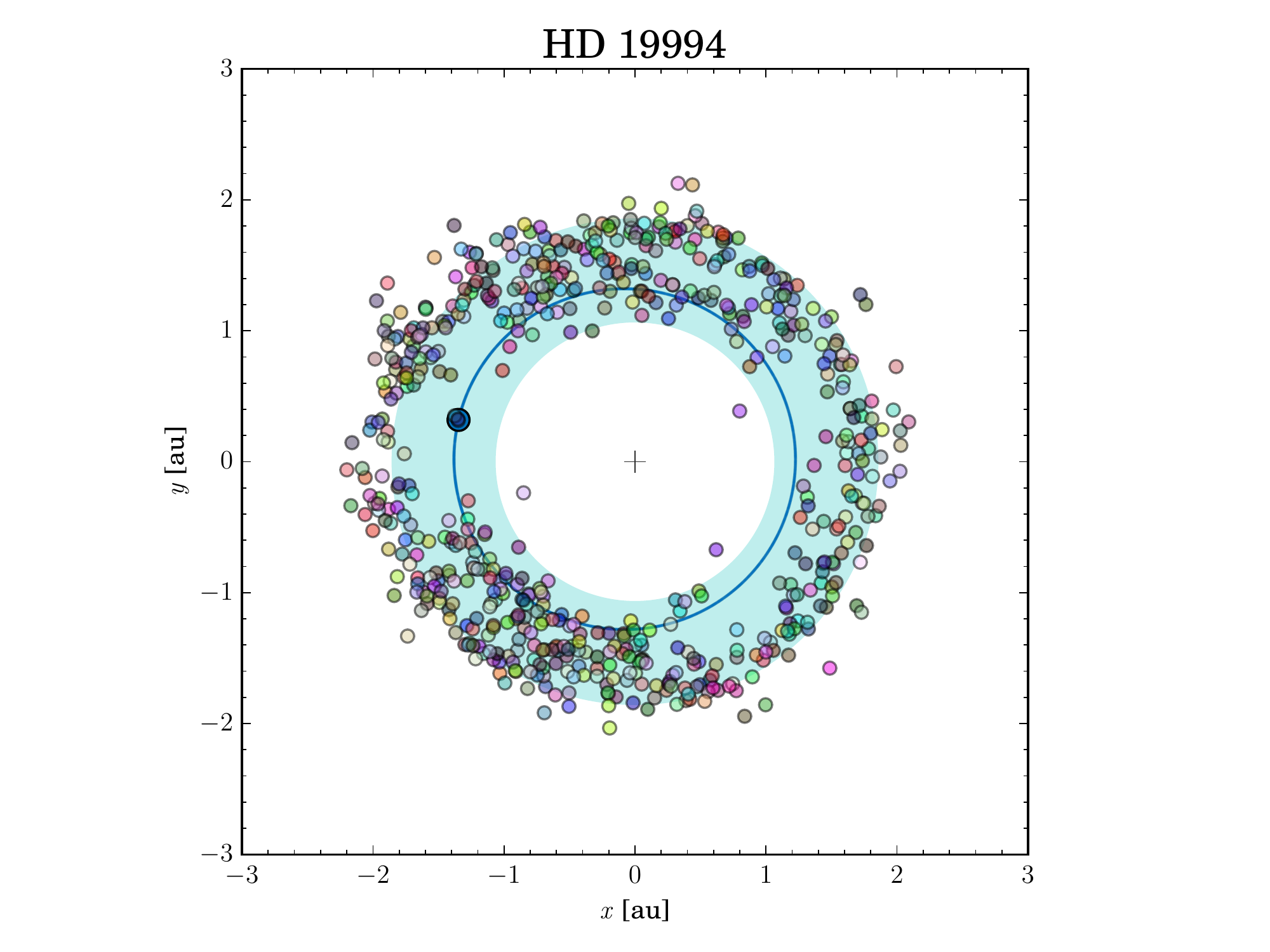}
        \caption{Cartesian Plot}\label{fig:mmr_1_cart}
	\end{subfigure}
	
    \begin{subfigure}{\linewidth}
	\centering
		\includegraphics[width=\linewidth]{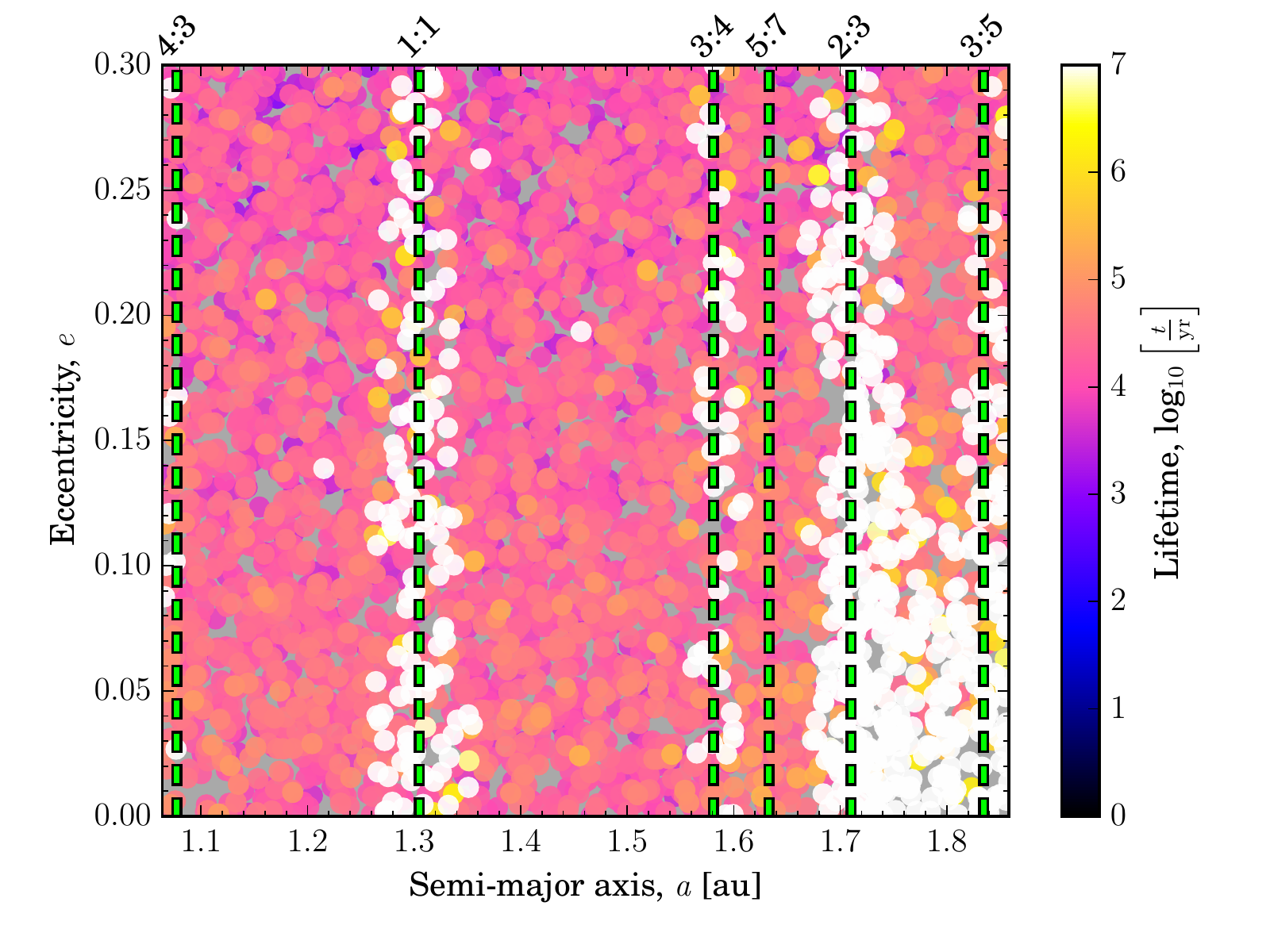}
		\caption{$a-e$ Map}\label{fig:mmr_1_init}
	\end{subfigure}
\caption{The HD 19994 system demonstrating the MMRs and L4 and L5 Lagrange points providing a means to stabilise an otherwise inherently unstable HZ. The Cartesian plot is a snapshot at the end of the simulation, with the existing Jovian planet's orbit shown in blue.}\label{fig:mmr_1}
\end{figure}
A Jovian planet whose orbit is embedded within the HZ would seem to suggest a completely unstable HZ. It is natural to suspect that, in such cases, the planet would not be able to coexist with a $1\ \textrm{M}_{\oplus}$ planet within the HZ. However, our simulations reveal several systems which demonstrate stability via mean-motion resonances (MMRs) with a HZ-embedded Jovian planet, including those in the form of planets trapped at the stable L4 and L5 Lagrange points, commonly referred to as Trojans. An example of such a class \rom{3} system is HD~19994 shown in Figure~\ref{fig:mmr_1}. Figure~\ref{fig:mmr_1_cart} shows the Cartesian view of the system at the conclusion of the simulation, clearly highlighting the stable Trojan companions that survive on the same orbit as the Jovian planet. We also see TPs that survive in the 2:3 and low eccentricity 3:5 MMRs. Figure~\ref{fig:mmr_1_init} shows the position of every particle tested on a semi-major axis versus eccentricity ($a-e$) parameter space, where the colour corresponds to the survival time of the particle on a logarithmic scale. This more clearly demonstrates the influence of stabilising MMRs (overlaid in green). The 1:1 and 2:3 MMRs offer particularly strong protection for TPs, but the influence of both the 3:4 and 3:5 MMRs can also be seen to result in a number of stable outcomes.

It should be kept in mind that these simulations used massless TPs, and so the mutual gravitational interactions between a possible $1\ \textrm{M}_{\oplus}$ planet and the Jovian planet have not been taken into account. However, based on the findings of \cite{Agnew2017}, it is often the case that $1\ \textrm{M}_{\oplus}$ planets are also able to survive in such simulations. Furthermore, the L4 and L5 Lagrangian points are stable for cases where the mass ratio of the Jovian planet to its host star is $\mu < 1/26$ which holds true for all the star--Jovian planet systems considered in this work \citep{Murray1999}.

An inherently challenging issue in the detection of planets that share an orbit with a Jupiter mass planet is that they represent a degenerate scenario for the radio velocity signal, and so would be indiscernible from the signal of a single Jupiter mass planet. This degeneracy would be broken if one (or both) planets transit or via differences in the planets long-term librations.

\subsubsection{Class~IV HZ--traversing Jovian planets}
\begin{figure}
	\begin{subfigure}{\linewidth}
    		\centering
    		\includegraphics[width=\linewidth]{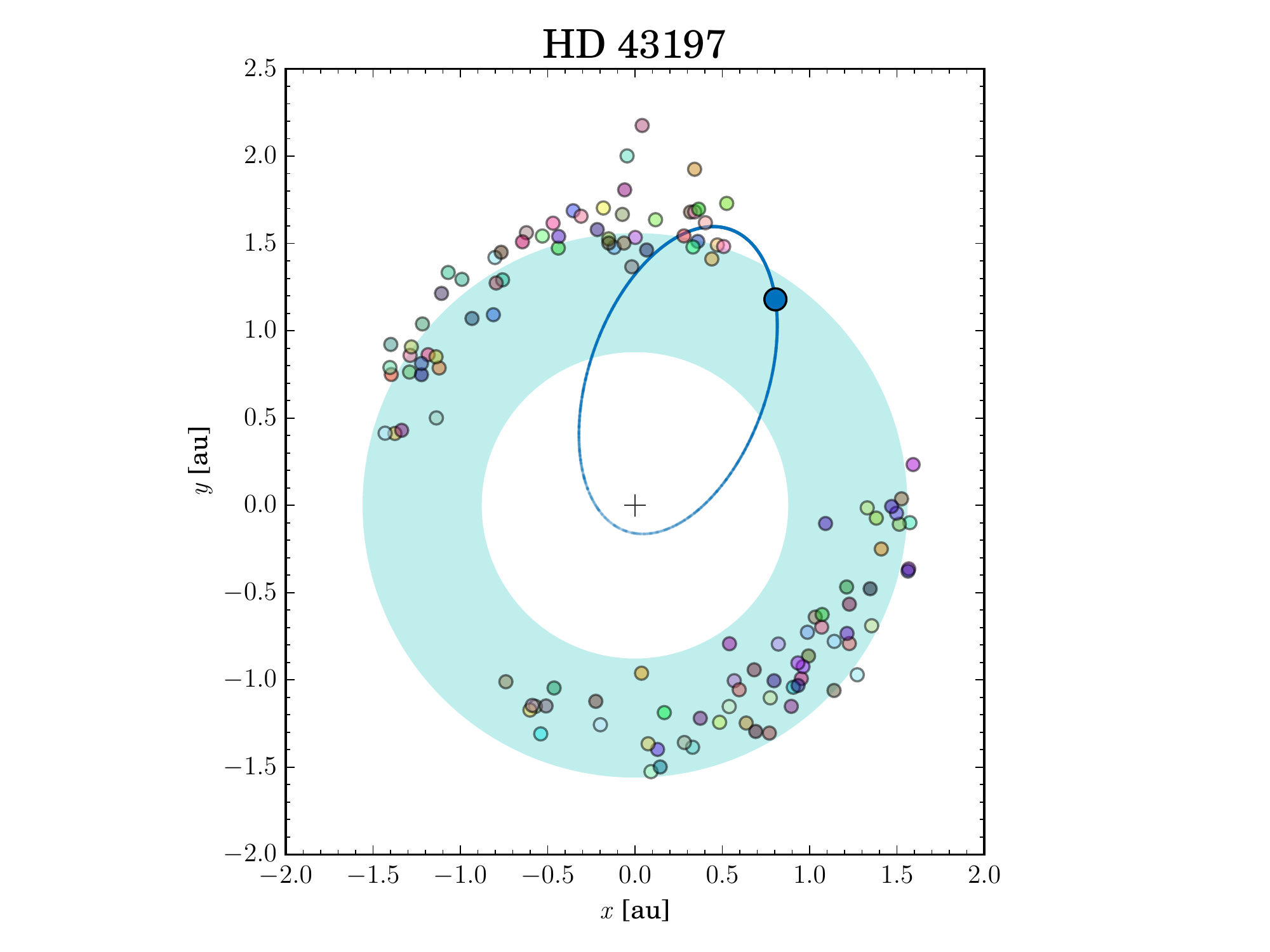}
        \caption{Cartesian}\label{fig:trav_cart}
	\end{subfigure}
	
    \begin{subfigure}{\linewidth}
	\centering
		\includegraphics[width=\linewidth]{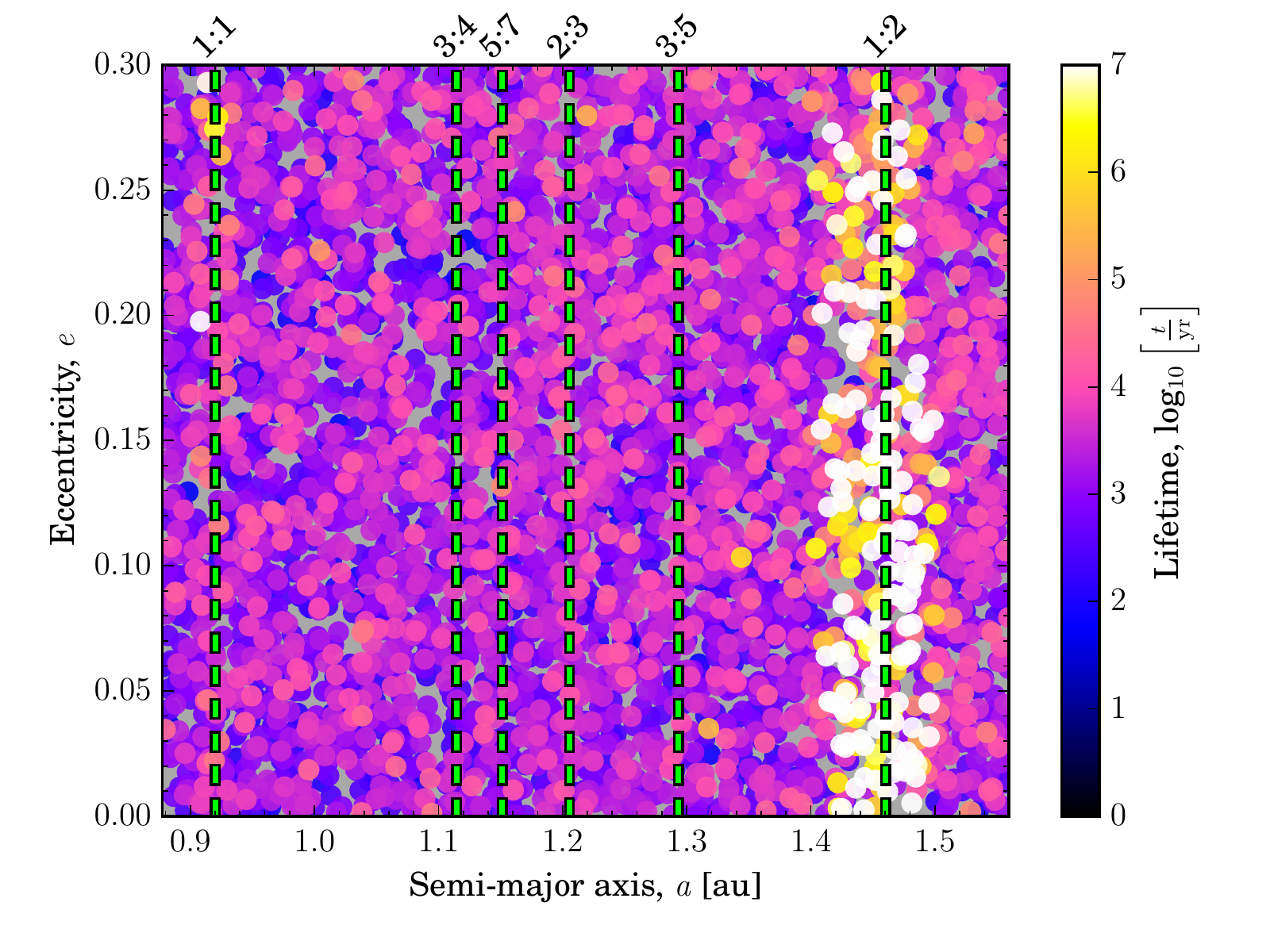}
		\caption{$a-e$ Map}\label{fig:trav_init}
	\end{subfigure}
\caption{The HD 43197 system demonstrating the 1:2 MMR providing a means to stabilise an otherwise inherently unstable HZ. The Cartesian plot is a snapshot at the end of the simulation, with the existing Jovian planet's orbit shown in blue.}\label{fig:trav_jup}
\end{figure}
Due to the higher eccentricities, a planetary architecture with a more unstable HZ is likely that of a HZ-traversing Jovian planet. As was the case with the example shown in section \ref{subsubsec:mmr}, MMRs can again provide stability in such a scenario. Figure~\ref{fig:trav_jup} demonstrates one such system, HD~43197. We can see in Figure~\ref{fig:trav_cart} that there exist stable TPs at the conclusion of the simulation on orbits that straddle the HZ. Even though some of these particles may move on orbits that exit the HZ, it may still be possible for such high eccentricity planets ($e<0.7$) to remain habitable depending on the response rate of the atmosphere-ocean system \citep{Williams2002,Jones2005}. Figure~\ref{fig:trav_init} demonstrates that the source of stabilisation in this scenario is the 1:2 MMR with the existing Jovian planet.

While it is interesting to demonstrate that a HZ-traversing Jovian planet can coexist with bodies in the HZ of the system, it should be noted that a high eccentricity Jovian planet would most likely be the result of gravitational interactions with other massive bodies during the planetary system's evolution, and so it seems highly unlikely that a rocky planet could remain in the HZ after such dynamical interactions \citep{Carrera2016,Matsumura2016}. In addition, as a result of eccentricity harmonics and aliasing, a fraction of published eccentric single planets are actually multiple systems \citep{Anglada-escud2010,Anglada-Escude2010,Wittenmyer2013} and so such systems must be further examined to confirm they are high eccentricity single systems. The result nevertheless demonstrates that seemingly destructive systems are certainly capable of harbouring other bodies on stable orbits within the HZ through the influence of stabilising resonances.

\subsection{Population Properties}
We also search for correlations between the stability of the HZ and the observable system parameters. Since it is gravitational interactions that will determine the stability of the HZ, one might expect $M_\mathrm{pl}$, $e_\mathrm{pl}$, and $a_\mathrm{pl}$ to have an effect. Other observables include $T_\mathrm{eff}$ and stellar metallicity, for which we would not expect any correlation, though $T_\mathrm{eff}$ is correlated with $M_{\star}$, which likely affects the mass ratio.

We examine all the systems simulated and plot their semi-major axis, eccentricity, mass ratio, metallicity and effective temperature against one another in Figure~\ref{fig:all_trends}, with the colour of each point indicating the number of surviving TPs. Other than some very slight clustering with respect to mass ratio, $e_\mathrm{pl}$, and $a_\mathrm{pl}$, no clear trends are revealed by our analysis.  The clustering, however, does emphasise the expected dependence that mass ratio, semi-major axis and eccentricity have on the stability of the HZ. As such, we use the semi-analytic criterion from \cite{Giuppone2013} in order to introduce a parameter that incorporates these parameters. The equation they use for the reach of the chaotic region around a planet is 
\begin{align} 
\label{eqn:chaotic_region}
	\delta = C\mu^{2/7}a_{\textrm{pl}},
\end{align} 
where $C$ was calculated to be a constant equal to $1.57$ \citep{Duncan1989,Giuppone2013}, $\mu = M_\mathrm{pl}/M_{\star}$ is the mass ratio between the planet and its parent star, and $a_{\textrm{pl}}$ is the semi-major axis of the planet. While the equation was originally formulated by \cite{Wisdom1980} for circular orbits, \cite{Giuppone2013} mention that it offers an approximation for eccentric orbits (as an eccentric orbit will precess and sweep out the entire annulus bound by the apsides of the orbit). As such, we can calculate the chaotic region as
\begin{align} 
\label{eqn:crossing_orbit}
	a_{\textrm{pl}}(1-e) - \delta \leq \textrm{Chaotic Region} \leq a_{\textrm{pl}}(1+e) + \delta,
\end{align} 
where $e$ is the planet's eccentricity, and $\delta$ is defined as in equation \ref{eqn:chaotic_region}.
Hence, the width of the chaotic region can be calculated by
\begin{align} 
	a_\textrm{chaos} &= \left(a_\textrm{apastron} + \delta\right) - \left(a_\textrm{periastron} - \delta \right) \nonumber \\
					&= \left(a(1+e) + \delta\right) - \left(a(1-e) - \delta \right) \nonumber
\end{align} 
and substituting $\delta$ from equation \ref{eqn:chaotic_region} yields
\begin{align} \label{eqn:chaos_index}
	a_\textrm{chaos} &= \left(a(1+e) + C\mu^{2/7}a\right) - \left(a(1-e) - C\mu^{2/7}a \right), \nonumber \\
	\frac{a_\textrm{chaos}}{a} &= 2\left(e + C\mu^{2/7}\right) \, .
\end{align} 
This value, $a_\textrm{chaos}/a$, we refer to as the \textit{chaos value} for the following plots and discussion.

\begin{figure}
	\begin{subfigure}{\linewidth}
    		\centering
    		\includegraphics[width=\linewidth]{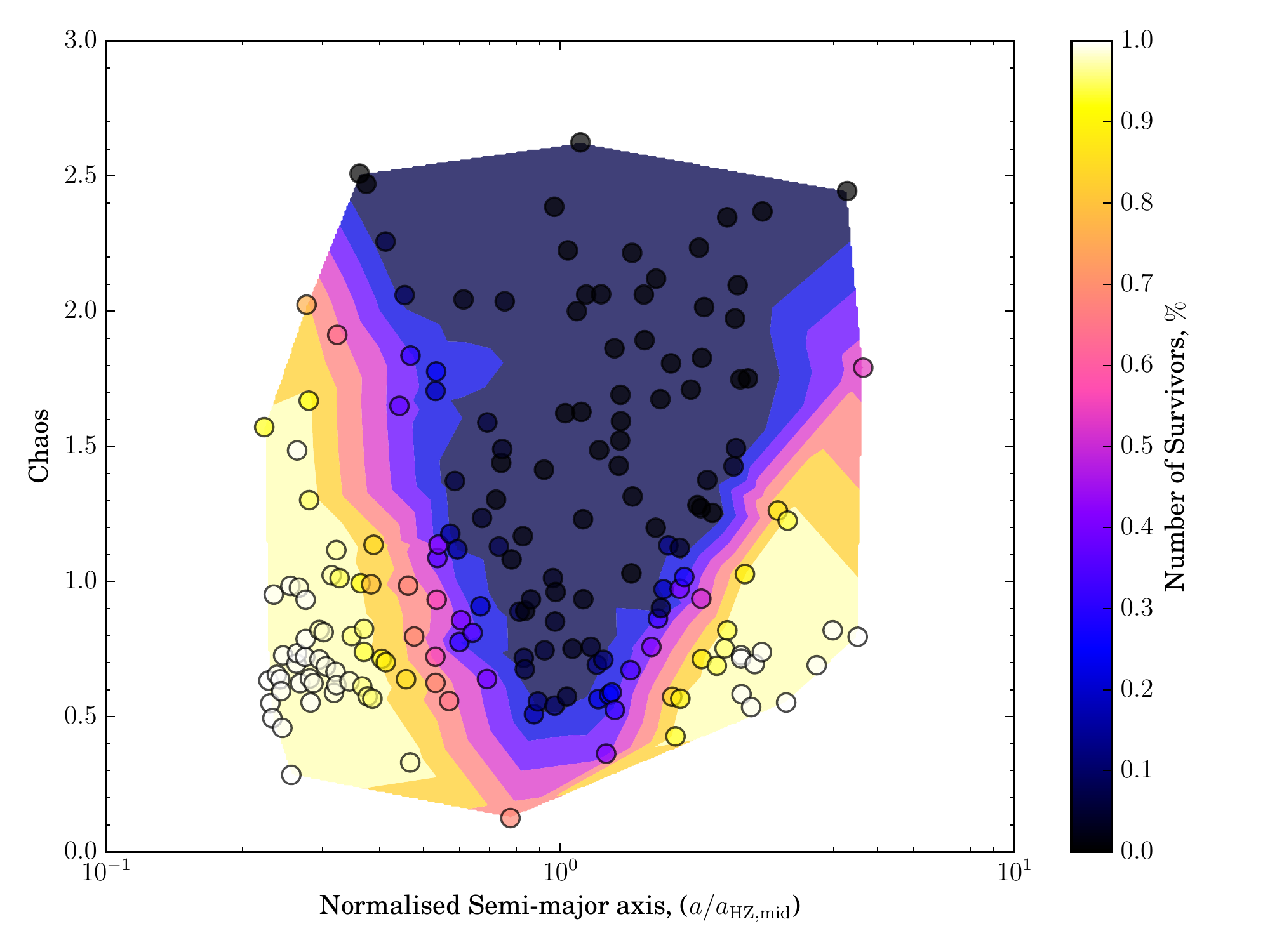}
        \caption{Individual Systems}\label{fig:contours_scatter}
	\end{subfigure}
	
    \begin{subfigure}{\linewidth}
	\centering
		\includegraphics[width=\linewidth]{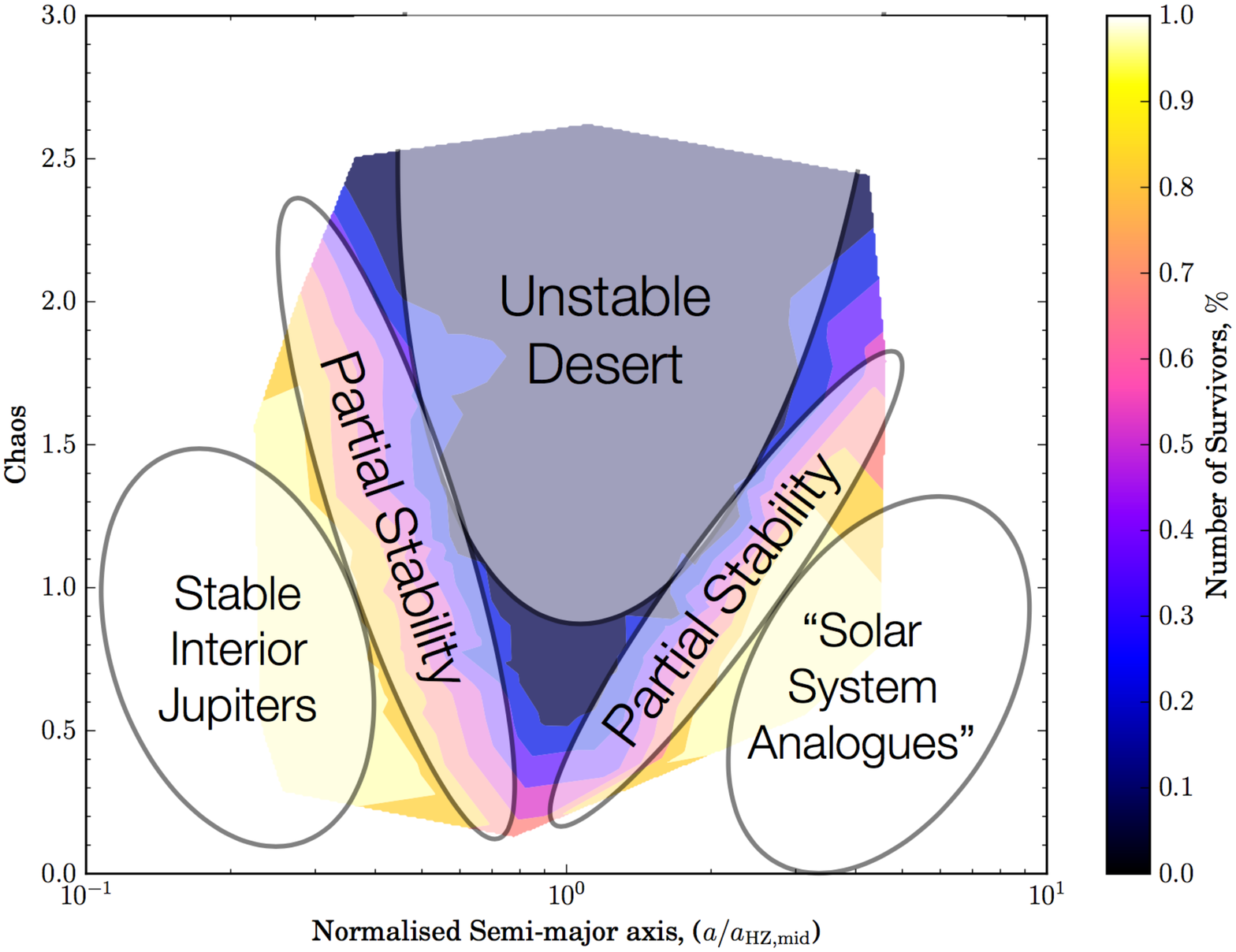}
		\caption{Classification Regions}\label{fig:contours_class}
	\end{subfigure}
\caption{The stability of all 182 single Jovian planet systems simulated in this study. The existing Jovian planet is plotted on a normalised semi-major axis ($a/a_\mathrm{HZ,mid}$) against chaos value (equation \ref{eqn:chaos_index}) parameter space. Fig~\ref{fig:contours_scatter} shows each individual system, where the colour of each point represents the percentage of surviving TPs at the end of the simulation. Contours have been interpolated and underlaid based on the 182 data points. Fig~\ref{fig:contours_class} divides the parameter space into different dynamical regions.}\label{fig:contours}
\end{figure}

In Figure~\ref{fig:contours} we plot all the systems by normalised semi-major axis ($a/a_\mathrm{HZ,mid}$) against the chaos value. It is immediately apparent that there is a very obvious ``desert'' where no TPs survive in the systems considered. As our simulations use massless TPs, this plot cannot be used to predict the stability of massive bodies, such as a $1\ \textrm{M}_{\oplus}$ planet. However, this plot can be used to exclude systems from further observational searches for planets in the HZ, since if TPs are dynamically unstable, then, in general, one would expect planets to also be dynamically unstable. However, there will be exceptions to this rule. Whilst adding an additional massive body (or, indeed, changing the mass of the giant planet in the system) will not affect the location of mean-motion resonances, the secular dynamics of the system will be impacted by such changes \citep{Raymond2006a,Raymond2005,Horner2008}. In rare cases, this might lead to an otherwise unstable orbit being stabilised. 

Newly discovered systems can be plotted on this map to predict whether it is worthwhile undertaking follow up observations, or exhaustive numerical simulations, to further investigate whether a $1\ \textrm{M}_{\oplus}$ planet could be hidden within the HZ on a stable orbit. As the number of planetary systems discovered continues to grow, having a quick method by which systems with unstable HZs can be removed from further studies will be beneficial. 

\subsection{Searching for Solar System Analogues}
\begin{figure}
	\centering
	\includegraphics[width=\linewidth]{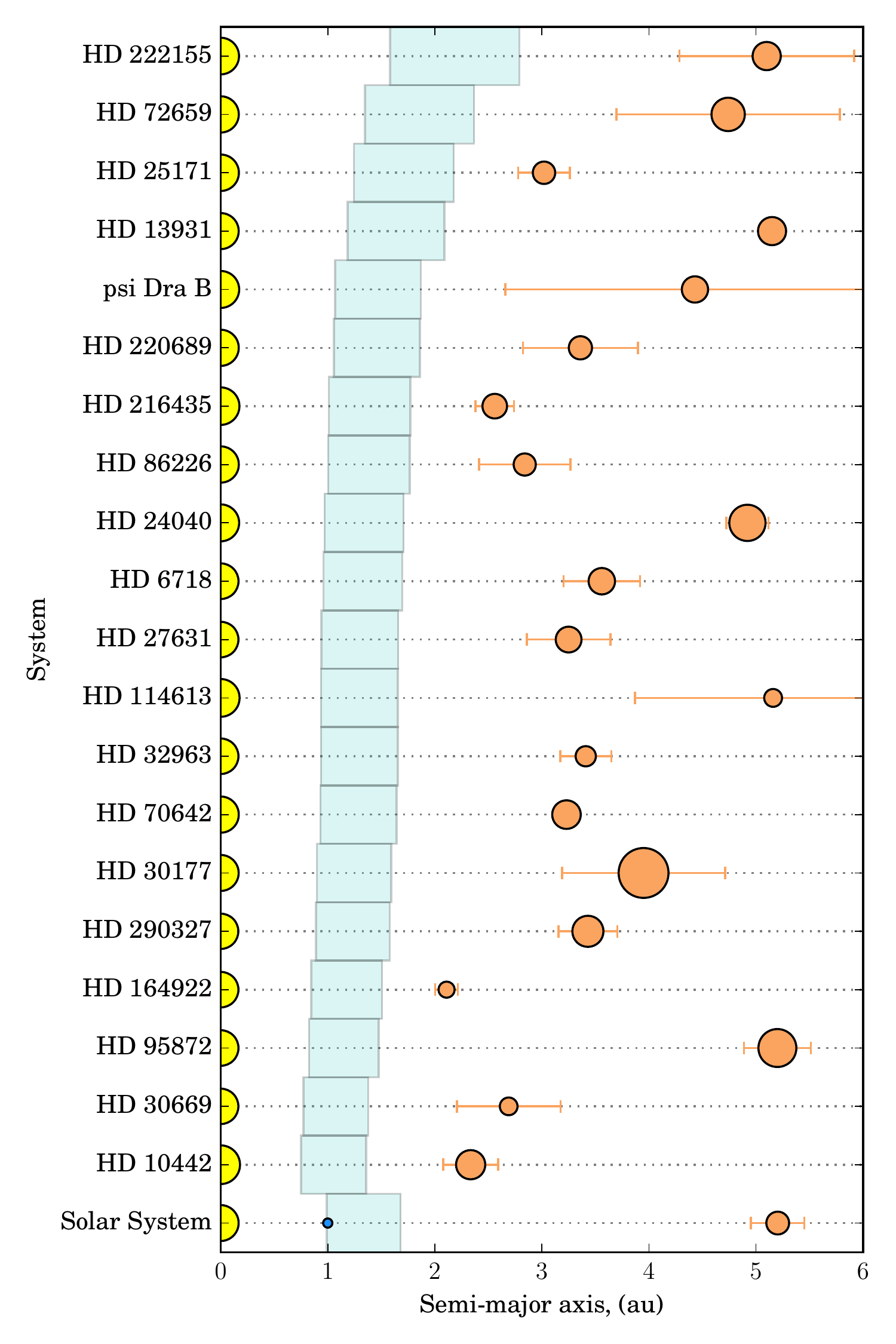}
	\caption{The 20 single Jovian planet systems that have been loosely classified as potential \textit{Solar system analogues}. This corresponds to a system that has a Jovian planet orbiting beyond the HZ, and a completely stable HZ within which a $1\ \textrm{M}_{\oplus}$ planet could be hidden. The error bars represent the apsides of the Jovian planet's orbit. The size of the point is proportional to $m^{1/3}$.}
	\label{fig:ss_analogs}
\end{figure}

From our investigation of the entire single Jovian population, we are able to provide a candidate list of potential \textit{Solar system analogues} for future planet hunters. These are shown with respect to the single Jovian planet population in Figure~\ref{fig:contours_class}, and schematically in Figure~\ref{fig:ss_analogs}. There are several ways in which a Solar system analogue can be defined. $\epsilon$ Eridani \citep{SchutzO2004,Backman2009,Greaves2014,Lestrade2015,MacGregor2015,Su2017} and HR 8799 \citep{Marois2008,Rhee2007,Su2009,Marois2010a,Matthews2014,Contro2016} are two such examples, both having multiple asteroid belts and hosting (or are proposed to host) several giant planets, just as we find in the Solar system. However, none have thus far been found to host rocky planets in their HZs. Another example is the recent discovery of the eighth planet in the Kepler-90 system \citep{Shallue2017}, just as we find 8 planets in our own system. However, these planets are all on far smaller, tighter orbits than planets in our Solar system. For this work, we use the term \textit{Solar system analogue} to encompass those systems that have a rocky Earth-like planet in the HZ, and a Jovian planet beyond the HZ \textit{with orbital periods similar to the Earth and Jupiter respectively}. Given that we use massless TPs in our simulations, we cannot constrain the stable semi-major axes of the HZ, as gravitational interactions between any putative exo-Earth and the Jovian planet are not taken into account. Thus we seek only those systems with a Jovian planet beyond the HZ that also have an entirely, or predominantly, unperturbed HZ. Figure~\ref{fig:ss_analogs} shows the 20 systems found in this study that fit this definition, with our Solar system (Earth and Jupiter only) for comparison.

Following \cite{Agnew2017}, we can compute the magnitude of the Doppler wobble that a $1\ \textrm{M}_{\oplus}$ planet would induce on its host star using the equation
\begin{align} 
\label{eqn:doppler_shift}
	K &= \left( \frac{2\pi G}{T_{\oplus}} \right)^{\frac{1}{3}} \frac{M_{\oplus}\sin{I}}{\left( M_\star + M_{\oplus} \right)^{\frac{2}{3}}} \frac{1}{\sqrt{1-e_{\oplus}^2}} \, ,
\end{align}  
where $G$ is the gravitational constant, $M_\star$ is the mass of the host star, $I$ is the inclination of the planet's orbit to our line of sight, and $T_{\oplus}$, $e_{\oplus}$ and $M_{\oplus}$ are the period, mass and eccentricity of the $1\ \textrm{M}_{\oplus}$ planet respectively. 

We next ask: what is the minimum radial velocity resolution required to detect an exo-Earth in the HZ if it exists? $K$ is larger for smaller orbital periods, so we calculate $K$ for the outer boundary of the HZ, which can be considered the ``conservative view'', i.e. the weakest Doppler wobble a rocky body would induce on its host star. Thus we calculate the semi-amplitude of the Doppler shift produced at the outer edge of the HZ to provide the minimum radial velocity resolution required to detect the exo-Earth \textit{if it exists}. These values are presented in Table~\ref{tab:rv_mins}. We also provide the resolution required for larger $2\ \textrm{M}_{\oplus}$ and $4\ \textrm{M}_{\oplus}$ planets, but acknowledge that the boundaries of the HZ will vary slightly for a larger planet, as noted by \cite{Kopparapu2014}. 

We also consider the dynamical evolution of the system, and what impact that may have had on the HZ. We use the definition of \textit{resilient habitability} introduced by \cite{Carrera2016}, which defines the ability of a planet to avoid being removed from a system (by collision or ejection) and remain within the HZ. \cite{Carrera2016} considered the dynamical interactions an existing Jovian planet would have with objects in the HZ during its evolution to the orbital parameters seen today. If a planet has resilient habitability, the HZ was not completely disturbed during the dynamical evolution of the system. \cite{Carrera2016} simulate a large suite of systems, and by scaling the results of these simulations to various different semi-major axes, create an $a-e$ map that can be used to infer the probability that a planet has resilient habitability, given the $a$ and $e$ values of an existing Jovian planet. We plot the Jovian planet from each of our Solar system analogues on the resilient habitability plots presented by \cite{Carrera2016}, and provide the probability bin each system falls into in Table~\ref{tab:rv_mins}. This provides another parameter by which to prioritise systems for observational follow-up to hunt for potentially habitable exo-Earths. Of the candidates we put forward, we are particularly interested in those that have a greater than $50\%$ probability of having resilient habitability.

We find that four systems have a resilient habitability probability of greater than $50\%$: HD 222155, HD 24040, HD 95872 and HD 13931, which has an almost $75\%$ probability. We suggest these should be the priority candidates for follow-up observation with ESPRESSO as they have not only dynamically stable HZs, but also have a greater than $50\%$ probability that the dynamical evolution of the Jovian planet did not completely destabilise the HZ. The candidate list presented in Table~\ref{tab:rv_mins} represents those systems we tested numerically to possess dynamically stable HZs. There are also those systems that we tagged as having a stable HZ (as discussed in section \ref{subsec:predicting_stable_regions}), 5 of which have a Jovian planet exterior to the HZ that may also be considered \textit{Solar system analogues}, however, we present only those systems for which numerical simulations were carried out. We also do not present those systems labelled as ``Stable Interior Jupiters'' in Figure~\ref{fig:contours_class}. These systems are shown to have very stable HZs. However, as they are interior to the HZ it raises another question regarding Jovian planet formation and migration scenarios, and what effect these may have on amount of material remaining for terrestrial planet formation \citep{Armitage2003,Mandell2003a,Fogg2005a,Mandell2007}.

As  discussed  in Section~\ref{subsec:dynamical_analysis}, our investigation and subsequent results are for perfectly co-planar systems. Mutually inclined planets may exchange angular momentum, resulting in the excitation of the orbital inclination and/or eccentricity of an otherwise potentially habitable world, which would clearly affect its HZ stability. As our Solar system analogues are defined such that the Jovian planet does not gravitationally perturb the HZ, it may also be the case that the Jovian planet is at a sufficient distance from the HZ that it would not strongly disturb the HZ at shallow mutual inclinations either. Regardless, we recommend further analysis of these systems of interest to more robustly prioritise them for observational follow-up.

\begin{table}
	\caption{The minimum required radial velocity sensitivities required to detect a $1\ \textrm{M}_{\oplus}$, $2\ \textrm{M}_{\oplus}$ or $4\ \textrm{M}_{\oplus}$ planet in the HZ. We also include the probability that each system has resilient habitability described by \protect\cite{Carrera2016}.}
   	\label{tab:rv_mins}
   	\centering
	\begin{tabular}{l c c c c}
        \toprule                   
        				& \multicolumn{3}{c}{$K_\mathrm{min}$ (m s$^{-1}$)} & Probability of \\
        	System		& $1\ \textrm{M}_{\oplus}$		& $2\ \textrm{M}_{\oplus}$	& $4\ \textrm{M}_{\oplus}$ & Resilient Habitability\\
        \midrule
        HD 222155	& 0.0504  & 0.1008   & 0.2016 & $50 - 75\%$\\   
        HD 72659 	& 0.0565  & 0.1129   & 0.2259 & $25 - 50\%$\\  
        HD 25171 	& 0.0581  & 0.1162   & 0.2323 & $25 - 50\%$\\   
        HD 13931		& 0.0613  & 0.1226   & 0.2452 & $50 - 75\%$\\
        psi Dra B 	& 0.0600  & 0.1200   & 0.2400 & $0 - 25\%$\\
        HD 220689	& 0.0643  & 0.1286   & 0.2573 & $25 - 50\%$\\
        HD 216435	& 0.0589  & 0.1179   & 0.2357 & $0 - 25\%$\\
        HD 86226 	& 0.0654  & 0.1308   & 0.2616 & $25 - 50\%$\\
        HD 24040 	& 0.0630  & 0.1260   & 0.2520 & $50 - 75\%$\\
        HD 6718 		& 0.0701  & 0.1402   & 0.2805 & $25 - 50\%$\\
        HD 27631 	& 0.0717  & 0.1433   & 0.2866 & $25 - 50\%$\\
        HD 114613 	& 0.0596  & 0.1193   & 0.2385 & $25 - 50\%$\\
        HD 32963		& 0.0717  & 0.1435   & 0.2870 & $25 - 50\%$\\
        HD 70642		& 0.0681  & 0.1361   & 0.2723 & $25 - 50\%$\\
        HD 30177		& 0.0705  & 0.1410   & 0.2820 & $25 - 50\%$\\
        HD 290327 	& 0.0751  & 0.1501   & 0.3002 & $25 - 50\%$\\
        HD 164922 	& 0.0752  & 0.1503   & 0.3006 & $0 - 25\%$\\
        HD 95872 	& 0.0755  & 0.1511   & 0.3022 & $50 - 75\%$\\
        HD 30669 	& 0.0795  & 0.1590   & 0.3179 & $0 - 25\%$\\
        HD 10442		& 0.0615  & 0.1229   & 0.2458 & $0 - 25\%$\\
        \toprule
   	\end{tabular}
\end{table}

\section{Conclusions}
\label{sec:summary}

We simulated the dynamical stability of the entire known single Jovian population for which stellar and planetary properties are available, that satisfy the criterion $2600$ K $\leq T_\mathrm{eff} \leq 7200$~K, and for which the Jovians are located within $0.1\ T_{\textrm{HZ}} \leq T_{\textrm{Jovian}} \leq 10\ T_{\textrm{HZ}}$.  We then investigated both the dynamical properties of individual systems as well as of the entire population with the aim of providing a guide for where to focus resources in the search for Earth-like planets and Solar system analogues.

We divide the 182 single Jovian systems into dynamical classes based on the apsides of the Jovian planet relative to the
boundaries of the HZ: systems for which the Jovian is interior to the HZ, exterior to the HZ, embedded in the HZ, and traverses the HZ. Perhaps somewhat surprisingly, we find that there are regions of stability in the HZ even when the Jovian is embedded in or traverses the HZ. For such dynamical classes, we find that stabilising MMRs are capable of providing regions of stability, including the L4 and L5 Lagrange points in the case of HZ-embedded Jovians. While these stable regions have been demonstrated with massless test particles in this study, \cite{Agnew2017} has shown that this is usually a strong indication that the region is also stable for a $1\ \textrm{M}_{\oplus}$. In the case of the L4 and L5 Lagrangian points, the regions are stable as long as the mass ratio of the Jovian planet is $\mu < 1/26$. It should be noted that these results are based on the systems \textit{as we see them today} and it has been shown that, particularly with the high eccentricity, HZ-traversing Jovian planets, the dynamical evolution may have already destroyed or ejected any bodies from the HZ \citep{Carrera2016,Matsumura2016}.

Examining the entire population, as expected we find that the main indicators of HZ stability are semi-major axis, mass ratio and eccentricity of the Jovian planet. Other observable quantities such as $T_{\textrm{eff}}$ and stellar metallicity show no overarching trends within the single Jovian planet population.  
However, by using the semi-analytic criterion of \cite{Giuppone2013}, we find a ``desert'' in the chaos--normalised semi-major axis parameter space which we can use to exclude single Jovian planet systems that are unable to coexist with other bodies in their HZ. This map of the \textit{chaos value} is useful for rapidly excluding newly discovered systems from further observational or numerical follow-up in the search for habitable exoplanets and Solar system analogues.

Systems with a completely stable HZ suggest that the Jovian planet's gravitational influence is not strong enough to interact with bodies within the HZ. In this work, we can simply define \textit{Solar system analogues} as systems with an exterior Jovian planet and an entirely, or predominantly, unperturbed HZ. We find that there are 20 systems which we can therefore define as Solar system analogues, with a Jovian planet exterior of the HZ, and for which the HZ is left completely, or nearly completely, unperturbed by its gravitational influence. These systems are ideal candidates to search for an Earth-like planet in the HZ of the system, and are shown in Figure~\ref{fig:ss_analogs} and Table~\ref{tab:rv_mins}. We also present the probability that each of these systems has resilient habitability as outlined by \cite{Carrera2016}. Specifically, we find that HD 222155, HD 24040, HD 95872 and HD 13931 all have a greater than $50\%$ probability (and in fact HD 13931 has an almost $75\%$ probability) that a terrestrial planet could  survive in the HZ during the dynamical evolution of the existing Jovian. As a result, we suggest that these systems should be a priority for observational follow-up with ESPRESSO.

While we were unable to find overarching properties amongst the single Jovian planet population that can be used to indicate HZ stability, we were able to generate a map over a parameter space which shows a clear clustering of systems with unstable HZs. This can be used to exclude newly discovered systems from an exhaustive suite of simulations or observational follow up in the search for habitable exoplanets. Furthermore, we have been able to provide a candidate list of potential Solar system analogues for use by planet hunters, which should assist in focusing resources in the search for habitable Earth-like planets amongst the ever-growing exoplanet population.

\section*{Acknowledgements}

We wish to thank the anonymous referee for helpful comments and suggestions that have improved the paper. MTA was supported by an Australian Postgraduate Award (APA). This work was performed on the gSTAR national facility at Swinburne University of Technology. gSTAR is funded by Swinburne and the Australian Government’s Education Investment Fund. This research has made use of the Exoplanet Orbit Database, the Exoplanet Data Explorer at exoplanets.org and the NASA Exoplanet Archive, which is operated by the California Institute of Technology, under contract with the National Aeronautics and Space Administration under the Exoplanet Exploration Program.




\bibliographystyle{mnras}
\bibliography{project_2.bib} 




\appendix
\section{Single Jovian Planet System Trends}
\begin{figure*}	
	\begin{subfigure}{\linewidth}
    		\centering
    		\includegraphics[width=\linewidth]{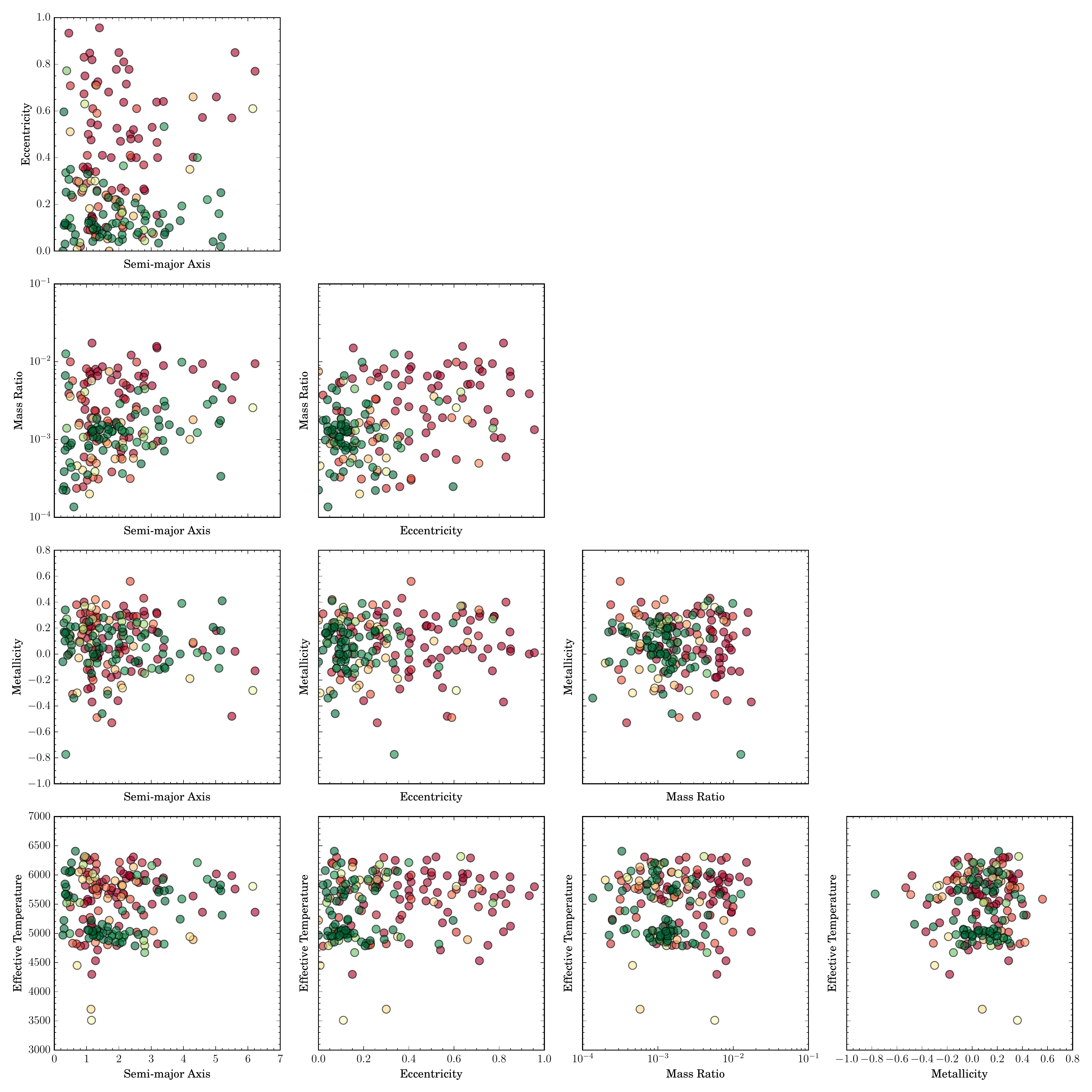}
	\end{subfigure}	
    \begin{subfigure}{\linewidth}
	\centering
		\includegraphics[width=\linewidth]{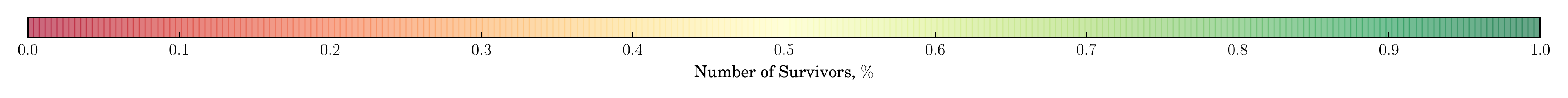}
	\end{subfigure}	
	\caption{Plots of all simulated single Jovian planet systems comparing the semi-major axes, eccentricities, mass ratios, metallicities and effective temperatures against one another. The colours indicate the number of survivors. As can be seen, other than some minor clustering with respect to semi-major axis, eccentricity and mass ratio, there is very little trending.}
	\label{fig:all_trends}
\end{figure*}


\bsp	
\label{lastpage}
\end{document}